\def\year{2018}\relax
\documentclass[letterpaper]{article} 
\usepackage{aaai18}  
\usepackage{times}  
\usepackage{helvet}  
\usepackage{courier}  
\usepackage{url}  
\usepackage{graphicx}  

\usepackage{amssymb}
 \usepackage{amsmath}
\usepackage{algorithm}
 \usepackage{algorithmic}
\usepackage{subfigure}
\usepackage{caption}
\usepackage{multirow}
\usepackage{float}
\usepackage{array}
\usepackage{epsfig}
\usepackage{xspace}
\usepackage{epstopdf}
\usepackage{latexsym}
\usepackage{bm,color,paralist,url}
\begin{document}
%
\title{A Survey on Network Embedding}
\author{Peng Cui$^{1}$, Xiao Wang$^{1}$, Jian Pei$^{2}$, Wenwu Zhu$^{1}$\\
$^{1}$Department of Computer Science and Technology, Tsinghua University, China\\
$^{2}$School of Computing Science, Simon Fraser University, Canada\\
cuip@tsinghua.edu.cn, wangxiao007@mail.tsinghua.edu.cn, \\jpei@cs.sfu.ca, wwzhu@tsinghua.edu.cn\\
}
\maketitle
\begin{abstract}
Network embedding assigns nodes in a network to
low-dimensional representations and effectively preserves the network structure. Recently,
a significant amount of progresses have been made toward this emerging network analysis paradigm. In
this survey, we focus on categorizing and then reviewing the current development on network embedding methods, and point out its future research directions.
We first summarize the motivation of network embedding. We discuss the classical graph embedding algorithms and
their relationship with network embedding. Afterwards and primarily, we provide a comprehensive overview
of a large number of network embedding methods in a systematic manner, covering the structure- and property-preserving network embedding methods, the network embedding methods with
side information and the advanced information preserving network embedding methods. Moreover, several evaluation approaches for network
embedding and some useful online resources, including the network data sets and softwares, are
reviewed, too.
Finally, we discuss the framework of exploiting these network
embedding methods to build an effective system and point out some potential future directions.
\end{abstract}

\section{Introduction}

Many complex systems take the form of networks, such as social networks, biological
networks, and information networks. It is well recognized that network data is often sophisticated and thus is challenging to deal with. To process network data effectively, the first critical challenge is to find effective network data representation, that is, how to represent networks concisely so that advanced analytic tasks, such as pattern discovery, analysis and prediction, can be conducted efficiently in both time and space.

Traditionally, we usually represent a network as a graph $G = \langle V, E\rangle$, where $V$ is a vertex set representing the nodes in a network, and $E$ is an edge set representing the relationships among the nodes. For large networks, such as those with billions of nodes, the traditional network representation poses several challenges to network processing and analysis.

\begin{itemize}
  \item \textbf{High computational complexity}. The nodes in a network are related to each other to a certain degree, encoded by the edge set $E$ in the traditional network representation. These relationships cause most of the network processing or analysis algorithms either iterative or combinatorial computation steps, which result in high computational complexity. For example, a popular way is to use the shortest or average path length between two nodes to represent their distance. To compute such a distance using the traditional network representation, we have to enumerate many possible paths between two nodes, which is in nature a combinatorial problem.  As another example, many studies assume that a node with links to important nodes tends to be important, and vice versa.  In order to evaluate the importance of a node using the traditional network representation, we have to iteratively conduct a stochastic node traversal process until reaching a convergence. Such methods using the traditional network representation result in high computational complexity that prevents them from being applicable to large-scale real-world networks.

  \item \textbf{Low parallelizability}. Parallel and distributed computing is de facto to process and analyze large-scale data. Network data represented in the traditional way, however, casts severe difficulties to design and implementation of parallel and distributed algorithms. The bottleneck is that nodes in a network are coupled to each other explicitly reflected by $E$.  Thus, distributing different nodes in different shards or servers often causes demandingly high communication cost among servers, and holds back speed-up ratio. Although some limited progress is made on graph parallelization by subtly segmenting large-scale graphs~\cite{staudtnetworkit}, the luck of these methods heavily depends on the topological characteristics of the underlying graphs. 

\item \textbf{Inapplicability of machine learning methods}. Recently, machine learning methods, especially deep learning, are very powerful in many areas. These methods provide standard, general and effective solutions to a broad range of problems. For network data represented in the traditional way, however, most of the off-the-shelf machine learning methods may not applicable. Those methods usually assume that data samples can be represented by independent vectors in a vector space, while the samples in network data (i.e., the nodes) are dependant to each other to some degree determined by $E$. Although we can simply represent a node by its corresponding row vector in the adjacency matrix of the network, the extremely high dimensionality of such a representation in a large graph with many nodes makes the in sequel network processing and analysis difficult.

\end{itemize}

The traditional network representation has become a bottleneck in large-scale network processing and analysis nowadays. Representing the relationships explicitly using a set of edges in the traditional representation is the upmost barrier.

\begin{figure}[t]
\centering
\subfigure[Input: karate network]{
\includegraphics[width=2in]{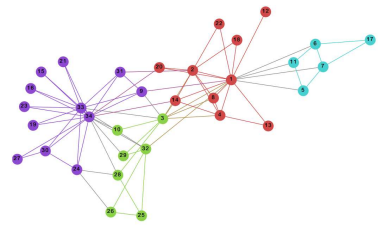}}
\subfigure[Output: representations]{
\includegraphics[width=2.5in]{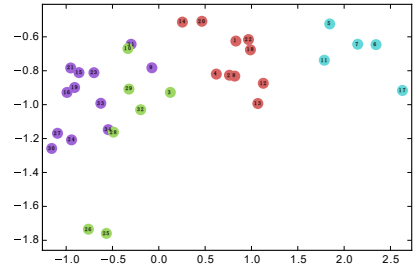}}
\caption{An example of network embedding on a karate network. Images are extracted from DeepWalk~\cite{perozzi2014deepwalk}.}
\label{ppp}
\end{figure}

To tackle the challenge, substantial effort has been committed to develop novel network embedding, i.e., learning low-dimensional vector
representations for network nodes. In the network embedding space, the relationships among the nodes, which were originally represented by edges or other high-order topological measures in graphs, is captured by the distances between nodes in the vector space, and the topological and structural characteristics of a node are encoded into its embedding vector. An example is shown in Fig.~\ref{ppp}.
After embedding the karate club network into a two-dimensional space, the similar nodes marked by the same color are
close to each other in the embedding space, demonstrating that the network structure can be well modeled in the two-dimensional embedding space.

\begin{figure*}
  \centering
  \includegraphics[width=6.50in]{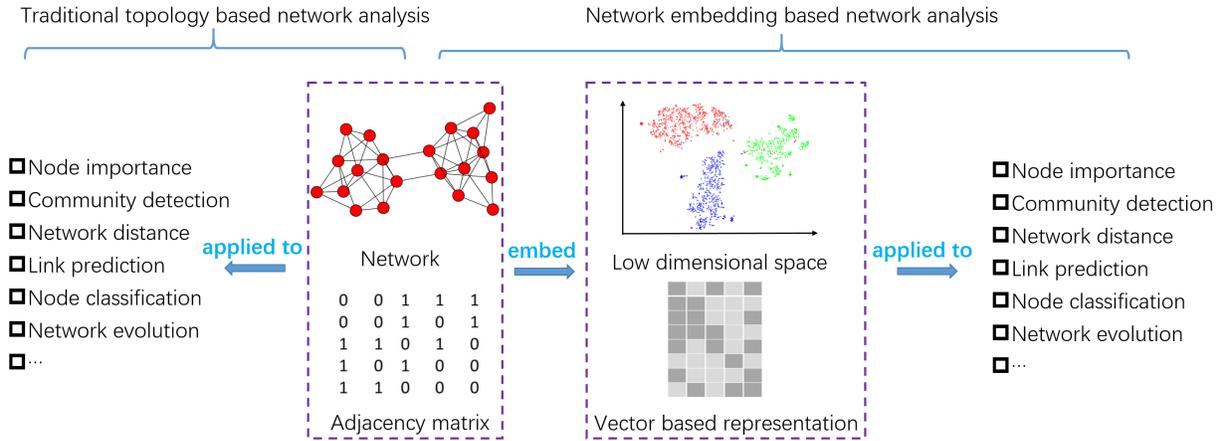}\\
  \caption{A comparison between network topology based network analysis and network embedding based network analysis. }\label{tra}
\end{figure*}

Network embedding, as a promising way of network representation, is capable of supporting subsequent network processing and analysis tasks such as node classification~\cite{sen2008collective,perozzi2014deepwalk}, node clustering~\cite{wang2017community}, network visualization~\cite{herman2000graph,wang2016kdd} and link prediction~\cite{liben2007link,ou2016kdd}. If this goal is fulfilled, the advantages of network embedding over traditional network representation methods are apparent, as shown in Fig.~\ref{tra}. The traditional
topology based network representation usually directly uses the observed adjacency matrix, which may contain noise or redundant information. The embedding based representation first aims to learn the dense and continuous representations of nodes in a low dimensional space, so that the noise or redundant information can be reduced and the intrinsic structure information can be preserved. As each node is represented by a vector containing its information of interest, many iterative or combinatorial problems in network analysis can be tackled by computing mapping functions, distance metrics or operations on the embedding vectors, and thus avoid high complexity. As the nodes are not coupling any more, it is convenient to apply main-stream parallel computing solutions for large-scale network analysis. Furthermore, network embedding can open the opportunities for network analysis to be benefited from the rich literature of machine learning. Many off-the-shelf machine learning methods such as deep learning models can be directly applied to solve network problems.

In order to make the embedding space well support network analysis tasks, there are two goals for network embedding. First, the original network can be reconstructed from the learned embedding space. It requires that, if there is an edge or relationship between two nodes, then the distance of these two nodes in the embedding space should be relatively small. In this way, the network relationships can be well preserved. Second, the learned embedding space can effectively
support network inference, such as predicting unseen links, identifying important nodes, and inferring node labels. It should be noted that an embedding space with only the goal of network reconstruction is not sufficient for network inference. Taking the link prediction problem as an example, if we only consider the goal of network reconstruction, the embedding vectors learned by SVD tend to fit all
the observed links and zero values in the adjacency matrix, which may lead to overfitting and cannot infer unseen links.

In this paper, we survey the state-of-the-art works on network embedding and point out future research directions.  In Section~\ref{sec:categorization}, we first categorize network embedding methods according to the types of information preserved in embedding, and summarize the commonly used models. We briefly review the traditional graph embedding methods and discuss the difference of these methods with the recent network embedding methods in Section~\ref{sec:graphEmbedding}.
Then, in Sections~\ref{sec:structure},~\ref{sec:sideinfo} and~\ref{sec:advanced}, we respectively review the methods on structure and property preserving network embedding, network embedding with side information, as well as advanced information preserving network embedding.
In Section~\ref{sec:app}, we present a few evaluation scenarios and some online resources, including the data sets and codes, for network embedding. We conclude and discuss a series of possible future directions
in Section~\ref{sec:future}.

\section{Categorization and The Models}\label{sec:categorization}

To support network inference, more information beyond nodes and links needs to be preserved in embedding space. Most research works on network embedding develop along this line in recent years. There are multiple ways to categorize them. In this paper, according to the types of information that are preserved in network embedding, we categorize the existing methods into three categories, that is, (1) network structure and properties preserving network embedding, (2) network embedding with side information and (3) advanced information preserving network embedding.

\begin{figure*}
  \centering
  \includegraphics[width=6.20in]{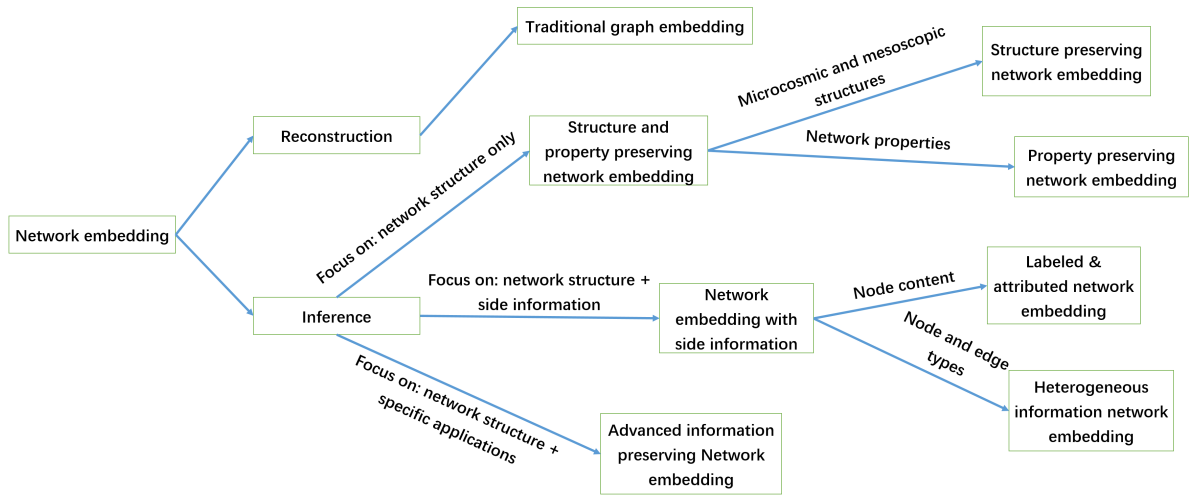}\\
  \caption{An overview of different settings of network embedding.}\label{overall1}
\end{figure*}

\subsection{The Categorization of Network Embedding Methods}

As mentioned before, network embedding usually has two goals, i.e., network
reconstruction and network inference. The traditional graph embedding methods, mainly focusing on network reconstruction, has been widely studied. We will briefly review those methods in Section~\ref{sec:graphEmbedding}. Fu and Ma~\cite{fu2012graph} present a more detailed survey. In this paper, we focus on the recently proposed network embedding methods aiming to address the goal of network inference. The categorization structure of the related works is shown in Fig.~\ref{overall1}.

\subsubsection{Structure and property preserving network embedding}

Among all the information encoded in a network, network structures and properties are two crucial factors that largely affect network inference. Consider a network with only topology information. Many network analysis tasks, such as identifying important nodes and predicting unseen links, can be conducted in the original network space. However, as mentioned before, directly conducting these tasks based on network topology has a series of problems, and thus poses a question that whether we can learn a network embedding space purely based on the network topology information, such that these tasks can be well supported in this low dimensional space. Motivated by this, attempts are proposed to preserve rich structural information into network embedding, from nodes and links~\cite{tang2015line} to neighborhood structure~\cite{perozzi2014deepwalk}, high-order proximities of nodes~\cite{wang2016kdd}, and community structures~\cite{wang2017community}. All these types of structural information have been demonstrated useful and necessary in various network analysis tasks. Besides this structural information, network properties in the original network space are not ignorable in modeling the formation and evolution of networks. To name a few, network transitivity (i.e. triangle closure) is the driving force of link formation in networks~\cite{huang2014mining}, and structural balance property plays an important role in the evolution of signed networks~\cite{cartwright1956structural}. Preserving these properties in a network embedding space is, however, challenging due to the inhomogeneity between the network space and the embedding vector space. Some recent studies begin to look into this problem and demonstrate the possibility of aligning these two spaces at the property level~\cite{ou2016kdd,wang2017signed}.

\subsubsection{Network Embedding with Side Information}

Besides network topology, some types of networks are accompanied with rich side information, such as node content or labels in information networks~\cite{tu2016max}, node and edge attributes in social networks~\cite{yang2015network}, as well as node types in heterogeneous networks~\cite{chang2015heterogeneous}. Side information provides useful clues for characterizing relationships among network nodes, and thus is helpful in learning embedding vector spaces. In the cases where the network topology is relatively sparse, the importance of the side information as complementary information sources is even more substantial. Methodologically, the main challenge is how to integrate and balance the topological and side information in network embedding. Some multimodal and multisource fusion techniques are explored in this line of research~\cite{natarajan2014inductive,yang2015network}.

\subsubsection{Advanced Information Preserving Network Embedding}

In the previous two categories, most methods learn network embedding in an unsupervised manner. That is, we only take the network structure, properties, and side information into account, and try to learn an embedding space to preserve the information. In this way, the learned embedding space is general and, hopefully, able to support various network applications. If we regard network embedding as a way of network representation learning, the formation of the representation space can be further optimized and confined towards different target problems. Realizing this idea leads to supervised or pseudo supervised information (i.e. the advanced information) in the target scenarios. Directly designing a framework of representation learning for a particular target scenario is also known as an end-to-end solution~\cite{li2017deepcas}, where high-quality supervised information is exploited to learn the latent representation space from scratch. End-to-end solutions have demonstrated their advantages in some fields, such as computer vision~\cite{yeung2016end} and natural language processing (NLP)~\cite{yang2017end}. Similar ideas are also feasible for network applications. Taking the network node classification problem as an example, if we have the labels of some network nodes, we can design a solution with network structure as input, node labels as supervised information, and embedding representation as latent middle layer, and the resulted network embedding is specific for node classification. Some recent works demonstrate the feasibility in applications such as cascading prediction~\cite{li2017deepcas}, anomaly detection~\cite{hu2016embedding}, network alignment~\cite{man2016predict} and collaboration prediction~\cite{chen2017task}.

\medskip
In general, network structures and properties are the fundamental factors that need to be considered in network embedding. Meanwhile, side information on nodes and links, as well as advanced information from target problem is helpful to enable the learned network embedding work well in real applications.

\subsection{ Commonly Used Models in Network Embedding}

To transform networks from original network space to embedding space, different models can be adopted to incorporate different types of information or address different goals. The commonly used models include matrix factorization, random walk, deep neural networks and their variations.

\subsubsection{Matrix Factorization}

An adjacency matrix is commonly used to represent the topology of a network, where each column and each row represent a node, and the matrix entries indicate the relationships among nodes. We can simply use a row vector or column vector as the vector representation of a node, but the formed representation space is $N$-dimensional, where $N$ is the total number of nodes. Network embedding, aiming to learn a low-dimensional vector space for a network, is eventually to find a low-rank space to represent a network, in contrast with the $N$-dimensional space. In this sense, matrix factorization methods, with the same goal of learning low-rank space for the original matrix, can naturally be applied to solve this problem. In the series of matrix factorization models, Singular Value Decomposition (SVD) is commonly used in network embedding due to its optimality for low-rank approximation~\cite{ou2016kdd}. Non-negative matrix factorization is often used because of its advantages as an additive model~\cite{wang2017community}.

\subsubsection{Random Walk}

As mentioned before, preserving network structure is a fundamental requirement for network embedding. Neighborhood structure, describing the local structural characteristics of a node, is important for network embedding. Although the adjacency vector of a node encodes the first-order neighborhood structure of a node, it is usually a sparse, discrete, and high-dimensional vector due to the nature of sparseness in large-scale networks. Such a representation is not friendly to subsequent applications. In the field of natural language processing (NLP), the word representation also suffers from similar drawbacks. The development of Word2Vector~\cite{mikolov2013distributed} significantly improves the effectiveness of word representation by transforming sparse, discrete and high-dimensional vectors into dense, continuous and low-dimensional vectors. The intuition of Word2Vector is that a word vector should be able to reconstruct the vectors of its neighborhood words which are defined by co-occurence rate. Some methods in network embedding borrow these ideas. The key problem is how to define ``neighborhood" in networks.

To make analogy with Word2Vector, random walk models are exploited to generate random paths over a network. By regarding a node as a word, we can regard a random path as a sentence, and the node neighborhood can be identified by co-occurence rate as in Word2Vector. Some representative methods include DeepWalk~\cite{perozzi2014deepwalk} and Node2Vec~\cite{jure2016kdd}.

\subsubsection{Deep Neural Networks}

By definition, network embedding is to transform the original network space into a low-dimensional vector space. The intrinsic problem is to learn a mapping function between these two spaces. Some methods, like matrix factorization, assume the mapping function to be linear. However, the formation process of a network is complicated and highly nonlinear, thus a linear function may not be adequate to map the original network to an embedding space.

If seeking for an effective non-linear function learning model, deep neural networks are certainly useful options because of their huge successes in other fields. The key challenges are how to make deep models fit network data, and how to impose network structure and property-level constraints on deep models. Some representative methods, such as SDNE~\cite{wang2016kdd}, SDAE~\cite{cao2016deep}, and SiNE~\cite{wang2017signed}, propose deep learning models for network embedding to address these challenges. At the same time, deep neural networks are also well known for their advantages in providing end-to-end solutions. Therefore, in the problems where advanced information is available, it is natural to exploit deep models to come up with an end-to-end network embedding solution. For instance, some deep model based end-to-end solutions are proposed for cascade prediction~\cite{li2017deepcas} and network alignment~\cite{man2016predict}.

\medskip
The network embedding models are not limited to those mentioned in this subsection. Moreover, the three kinds of models are not mutually exclusive, and their combinations are possible to make new solutions. More models and details will be discussed in later sections.

\section{Network Embedding v.s. Graph Embedding}\label{sec:graphEmbedding}

The goal of graph embedding is similar as network embedding, that is, to embed a graph into a low-dimensional vector space~\cite{yan2005graph}. There is a rich literature in graph embedding. Fu and Ma~\cite{fu2012graph} provide a thorough review on the traditional graph embedding methods. Here we only present some representative and classical methods on graph embedding, aiming to demonstrate the critical differences between graph embedding and the current network embedding.

\subsection{Representative Graph Embedding Methods}

Graph embedding methods are originally studied as dimension reduction techniques. A graph is usually constructed from a feature represented data set, like image data set. Isomap~\cite{tenenbaum2000global}
first constructs a neighborhood graph $G$ using connectivity algorithms such as $K$ nearest neighbors (KNN), i.e., connecting data entries $i$ and $j$ if $i$ is one of the $K$ nearest neighbors of
$j$. Then based on $G$, the shortest path $d_{ij}^G$ of entries $i$ and $j$ in $G$ can be computed. Consequently, for all the $N$ data entries in the data set,
we have the matrix of graph distances $D_G=\{d_{ij}^G\}$. Finally, the classical multidimensional scaling (MDS) method is applied to
$D_G$ to obtain the coordinate vector $\mathbf{u}_i$ for entry $i$, which aims to minimize the following function:
\begin{equation}
\label{isomap}
\sum_{i=1}^N\sum_{j=1}^N(d_{ij}^G-\|\mathbf{u}_i-\mathbf{u}_j\|)^2.
\end{equation}
Indeed, Isomap learns the representation $\mathbf{u}_i$ of entry $i$, which approximately preserves the geodesic distances of the entry pairs in the low-dimensional space.

The key problem of Isomap is its high complexity due to the computing of pair-wise shortest pathes.
Locally linear embedding (LLE)~\cite{roweis2000nonlinear} is proposed to eliminate the need to estimate the pairwise distances between widely separated entries.
LLE assumes that each entry and its neighbors lie on or close to a locally linear patch of a mainfold. To characterize
the local geometry, each entry can be reconstructed from its neighbors as follows:
\begin{equation}
\label{lle1}
\min_{\mathbf{W}}\sum_i\|\mathbf{x}_i-\sum_jW_{ij}\mathbf{x}_j\|^2,
\end{equation}
where the weight $W_{ij}$ measures the contribution of the entry $\mathbf{x}_j$ to the reconstruction of entry $\mathbf{x}_i$.
Finally, in the low-dimensional space, LLE constructs a neighborhood-preserving mapping based on
locally linear reconstruction as follows:
\begin{equation}\label{lle2}
\min_{\mathbf{U}}\sum_{i}\|\mathbf{u}_i-\sum_jW_{ij}\mathbf{u}_j\|^2.
\end{equation}
By optimizing the above function, the low-dimensional representation matrix $\mathbf{U}$, which preserves
the neighborhood structure, can be obtained.

Laplacian eigenmaps (LE)~\cite{belkin2002laplacian} also begins with constructing a graph using $\epsilon$-neighborhoods or $K$
nearest neighbors. Then the heat kernel~\cite{berline2003heat} is utilized to choose the weight $W_{ij}$ of nodes $i$ and $j$
in the graph. Finally, the representation $\mathbf{u}_i$ of node $i$ can be obtained by minimizing the following function:
\begin{equation}\label{le}
\sum_{i,j}\|\mathbf{u}_i-\mathbf{u}_j\|^2W_{ij}=tr(\mathbf{U}^T\mathbf{LU}),
\end{equation}
where $\mathbf{L}=\mathbf{D}-\mathbf{W}$ is the Laplacian matrix, and $\mathbf{D}$ is the diagonal matrix with $D_{ii}=\sum_jW_{ji}$.
In addition, the constraint $\mathbf{U}^T\mathbf{DU}=\mathbf{I}$ is introduced to avoid trivial solutions.
Furthermore, the locality preserving projection (LPP)~\cite{he2004locality}, a linear approximation of the nonlinear LE, is proposed. Also, it introduces a transformation matrix $\mathbf{A}$ such that the representation $\mathbf{u}_i$ of entry $\mathbf{x}_i$ is $\mathbf{u}_i=\mathbf{A}^T\mathbf{x}_i$.
LPP computes the transformation matrix
$\mathbf{A}$ first, and finally the representation $\mathbf{u}_i$ can be obtained.

\begin{figure*}
  \centering
  \includegraphics[width=6.30in]{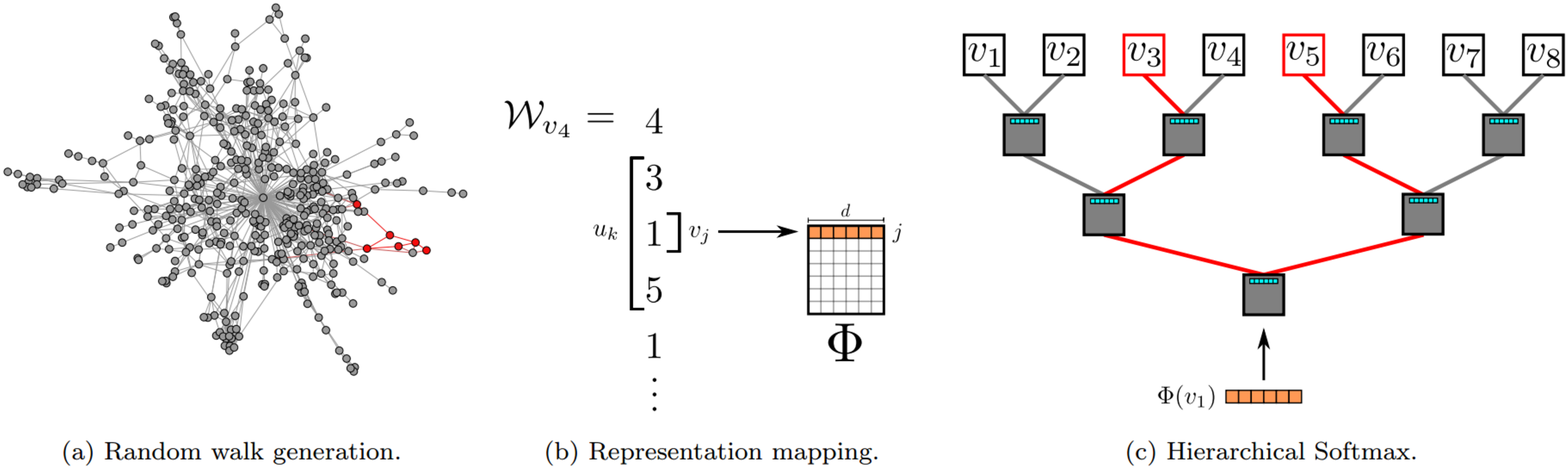}\\
  \caption{Overview of DeepWalk. Image extracted from~\cite{perozzi2014deepwalk}. }\label{dww}
\end{figure*}

These methods are extended in the rich literature of graph embedding by considering different characteristics of the constructed graphs~\cite{fu2012graph}.

\subsection{Major Differences}

Network embedding and graph embedding have substantial differences in objective and assumptions. As mentioned before, network embedding has two goals, i.e. reconstructing original networks and support network inference. The objective functions of graph embedding methods mainly target the goal of graph reconstruction. As discussed before, the embedding space learned for network reconstruction is not necessarily good for network inference. Therefore, graph embedding can be regarded as a special case of network embedding, and the recent research progress on network embedding pays more attention to network inference.

Moreover, graph embedding mostly works on graphs constructed from feature represented data sets, where the proximity among nodes encoded by the edge weights are well defined in the original feature space. In contrast, network embedding mostly works on naturally formed networks, such as social networks, biology networks, and e-commerce networks. In those networks, the proximities among nodes are not explicitly or directly defined. The definition of node proximities depends on specific analytic tasks and application scenarios. Therefore, we have to incorporate rich information, such as network structures, properties, side information and advanced information, in network embedding to facilitate different problems and applications.

In the rest of the paper, we mainly focus on the network embedding methods with the goal of supporting network inference.

\section{Structure and Property Preserving Network Embedding}
\label{sec:structure}

In essence, one basic requirement
of network embedding is to appropriately preserve network structures and capture properties of networks.
Often, network structures include first-order structure and higher-order structure, such as second-order
structure and community structure. Networks with different types have different properties. For example, directed networks have the asymmetric transitivity property. The structural balance theory is widely applicable to signed networks.

In this section, we review the representative methods of structure preserving network
embedding and property preserving network embedding.

\subsection{Structure Preserving Network Embedding}

Network structures can be categorized into different groups that present at different granularities. The commonly exploited network structures in network embedding include neighborhood structure, high-order node proximity and network communities.

\subsubsection{Neighborhood Structures and High-order Node Proximity}

DeepWalk~\cite{perozzi2014deepwalk} is proposed for learning the representations of nodes in a network, which
is able to preserve the neighbor structures of nodes. DeepWalk discovers that the distribution of nodes appearing
in short random walks is similar to the distribution of words in natural language. Motivated by this observation, Skip-Gram model~\cite{mikolov2013distributed}, a widely
used word representation learning model, is adopted by DeepWalk to
learn the representations of nodes. Specifically, as shown in Fig.~\ref{dww}, DeepWalk adopts a truncated random walk on a network to generate a set of walk
sequences. For each walk sequence $s=\{v_1,v_2,...,v_s\}$, following Skip-Gram, DeepWalk aims to maximize the probability
of the neighbors of node $v_i$ in this walk sequence as follows:\begin{equation}\label{dw}
  \max_{\phi} Pr(\{v_{i-w},...,v_{i+w}\}\backslash v_i|\phi (v_i))=\Pi_{j=i-w,j\neq i}^{i+w}Pr(v_j|\phi (v_i)),
\end{equation}
where $w$ is the window size, $\phi (v_i)$ is the current representation of $v_i$ and $\{v_{i-w},...,v_{i+w}\}\backslash v_i$ is
the local context nodes of $v_i$. Finally, hierarchical soft-max~\cite{mikolov2013efficient} is used to efficiently infer the embeddings.

Node2vec~\cite{jure2016kdd} demonstrates that DeepWalk is not expressive enough to capture the diversity of connectivity patterns in a network. Node2vec defines a flexible notion of a node's network neighborhood and designs a second order
random walk strategy to sample the neighborhood nodes, which can smoothly interpolate between breadth-first
 sampling (BFS) and depth-first sampling (DFS). Node2vec is able to learn the representations
that embed nodes with same network community closely, and to learn representations where nodes sharing similar
roles have similar embeddings.

\begin{figure}
  \centering
  \includegraphics[width=2.50in]{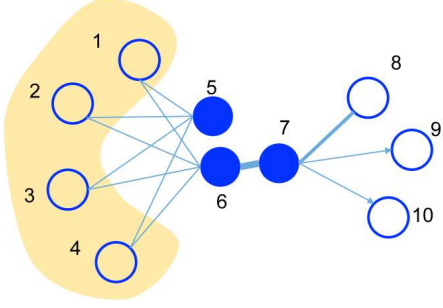}\\
  \caption{An example of the first-order and second-order structures in a network. Image extracted from~\cite{tang2015line}. }\label{line}
\end{figure}

LINE~\cite{tang2015line} is proposed for large scale network embedding, and can preserve the first and second order proximities.
The first order proximity is the observed pairwise proximity between two nodes, such as the observed edge between nodes 6 and 7 in Fig.~\ref{line}. The second order proximity is determined by
the similarity of the ``contexts" (neighbors) of two nodes. For example, the second order similarity between nodes 5 and 6 can be obtained by their neighborhoods $1$, $2$, $3$, and $4$ in Fig.~\ref{line}. Both the first order and second order proximities are important in measuring the relationships between two nodes.
The first order proximity can be measured by the joint probability distribution between two nodes $v_i$ and $v_j$ as
\begin{equation}\label{line1}
  p_1(v_i,v_j)=\frac{1}{1+exp(-\mathbf{u}_i^T\mathbf{u}_j)}.
\end{equation}
The second order proximity is modeled by the probability of the context node $v_j$ being generated by node $v_i$, that is,
\begin{equation}\label{line2}
  p_2(v_j|v_i)=\frac{exp(\bar{\mathbf{u}}_j^T\bar{\mathbf{u}}_i)}{\sum_k exp(\bar{\mathbf{u}}_k^T\bar{\mathbf{u}}_i))}.
\end{equation}
The conditional distribution implies that nodes with similar distributions over the contexts are similar to each other.
By minimizing the KL-divergence of the two distributions and the empirical distributions respectively, the representations of nodes
that are able to preserve the first and second order proximities can be obtained.

Considering that LINE only preserves the first-order and second-order proximities, GraRep~\cite{cao2015grarep} demonstrates that $k$-step ($k>2$) proximities
should also be captured when constructing the global representations of nodes. Given the adjacency matrix $\mathbf{A}$, the $k$-step probability
transition matrix can be computed by $\mathbf{A}^k=\underbrace{\mathbf{A}...\mathbf{A}}_k$, whose element
$A^k_{ij}$ refers to the transition probability $p_k(j|i)$ from a current node $i$ to a context node $j$ and the
transition consists of $k$ steps. Moreover, motivated by the Skip-Gram model~\cite{mikolov2013distributed}, the
$k$-step loss function of node $i$ is defined as
\begin{equation}\label{grarep}
  L_k(i)=(\sum_j p_k(j|i)\log \sigma (\mathbf{u}_j^T\mathbf{u}_i))+\lambda \mathbb{E}_{j'\sim p_k(V)}[\log \sigma (-\mathbf{u}_i^T\mathbf{u}_{j'})],
\end{equation}
where $\sigma (x)=(1+e^{-x})^{-1}$, $p_k(V)$ is the distribution over the nodes in the network and $j'$ is the node
obtained from negative sampling. Furthermore, GraRep reformulates the loss function as the matrix factorization problem,
for each $k$-step loss function, SVD can be directly
used to infer the representations of nodes. By concentrating the representations learned from each function, the global
representations can be obtained.

\subsubsection{Network Communities}

Wang~\textit{et~al.}~\cite{wang2017community} propose a modularized nonnegative matrix factorization (M-NMF) model for network
embedding, which aims to preserve both the microscopic structure, i.e., the first-order and second-order proximities
of nodes, and the mesoscopic community structure~\cite{girvan2002community}. To preserve the microscopic structure,
they adopt the NMF model~\cite{lee2001algorithms} to factorize the pairwise node similarity matrix and learn the representations of nodes.
Meanwhile, the community structure is detected by modularity maximization~\cite{newman2006finding}. Then, based on
the assumption that if the representation of a node is similar to that of a community, the node may have a
high propensity to be in this community, they introduce
an auxiliary community representation matrix to bridge the representations of nodes with the community structure.
In this way, the learned representations of nodes are constrained by both the microscopic structure and community structure, which
contains more structural information and becomes more discriminative.

\begin{figure}
  \centering
  \includegraphics[width=3.20in]{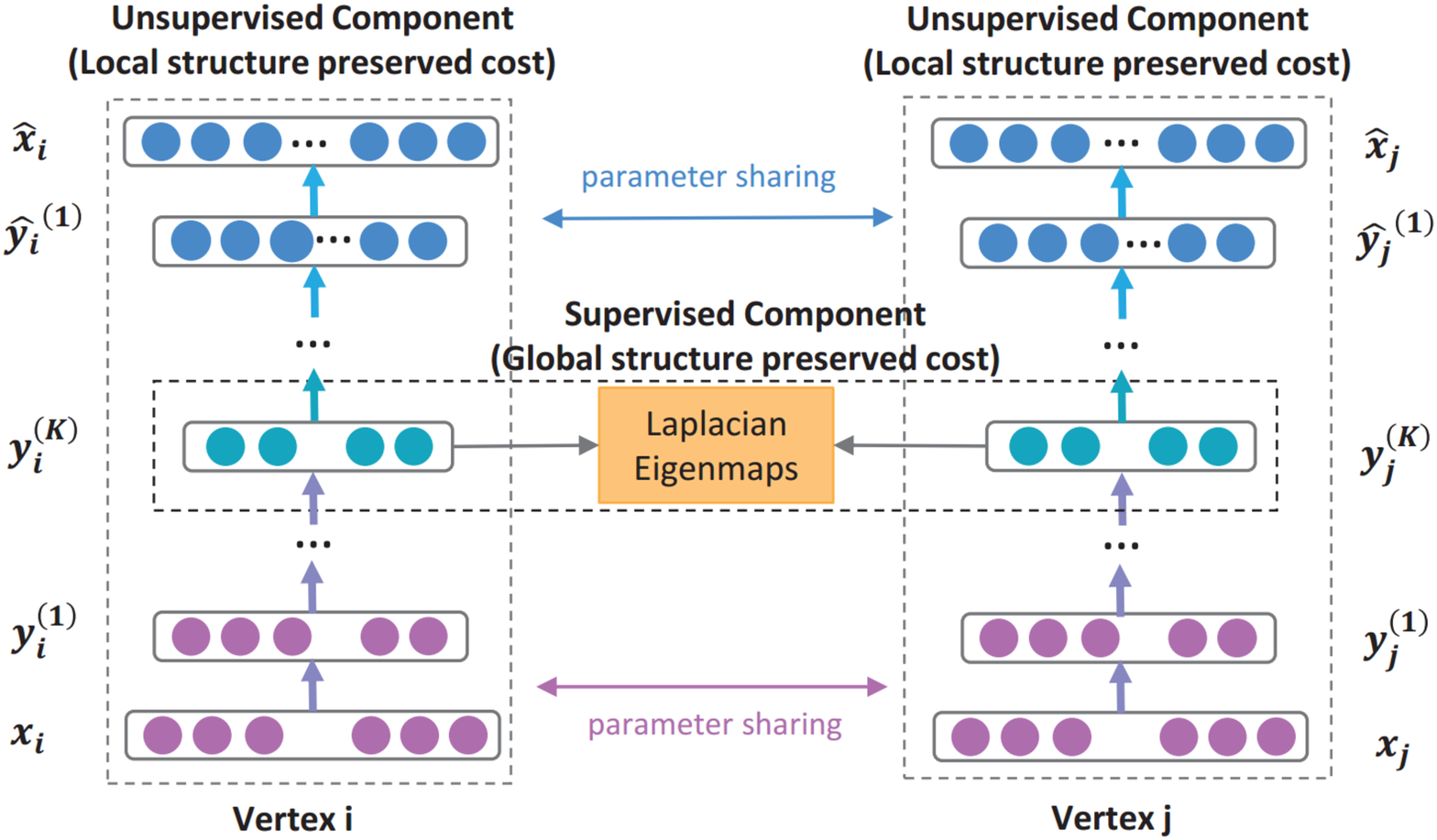}\\
  \caption{The framework of SDNE. Image extracted from~\cite{wang2016kdd}. }\label{sdne}
\end{figure}

The aforementioned methods mainly adopt the shallow models, consequently, the representation ability is limited. SDNE~\cite{wang2016kdd} proposes a deep model for network embedding, so as to address the high non-linearity, structure-preserving, and sparsity issues. The framework is shown in Fig.~\ref{sdne}. Specifically, SDNE uses the deep autoencoder with multiple non-linear layers to preserve the neighbor structures of nodes. Given the input adjacency nodes $\mathbf{x}_i$ of node $i$, the hidden representations for each layer can be obtained by
\begin{equation}\label{sdne1}
\begin{split}
  &\mathbf{y}_i^{(1)}=\sigma (\mathbf{W}^{(1)}\mathbf{x}_i+\mathbf{b}^{(1)})\\
  &\mathbf{y}_i^{(k)}=\sigma (\mathbf{W}^{(k)}\mathbf{y}_i^{(k-1)}+\mathbf{b}^{(k)}),k=2,...,K.
\end{split}
\end{equation}
Then the output representation $\mathbf{\hat{x}}_i$ can be obtained by reversing the calculation process of encoder.
To impose more penalty to the reconstruction error of the non-zero elements than that of zero elements, SDNE introduces the penalty vector
$\mathbf{b}_i=\{b_{ij}\}_{j=1}^n$ ($b_{ij}$ is larger than a threshold if there is an edge between nodes $i$ and $j$) and gives rise to the following function that can preserve the second-order proximity
\begin{equation}\label{sdne2}
  \mathcal{L}_{2nd}=\sum_i\|(\mathbf{\hat{x}}_i-\mathbf{x}_i)\odot \mathbf{b}_i\|^2.
\end{equation}
To preserve the first-order proximity of nodes, the idea of Laplacian eigenmaps~\cite{belkin2002laplacian} is adopted.
By exploiting the first-order and second-order proximities jointly into the learning process, the representations of nodes
can be finally obtained.

\begin{figure*}
  \centering
  \includegraphics[width=6.50in]{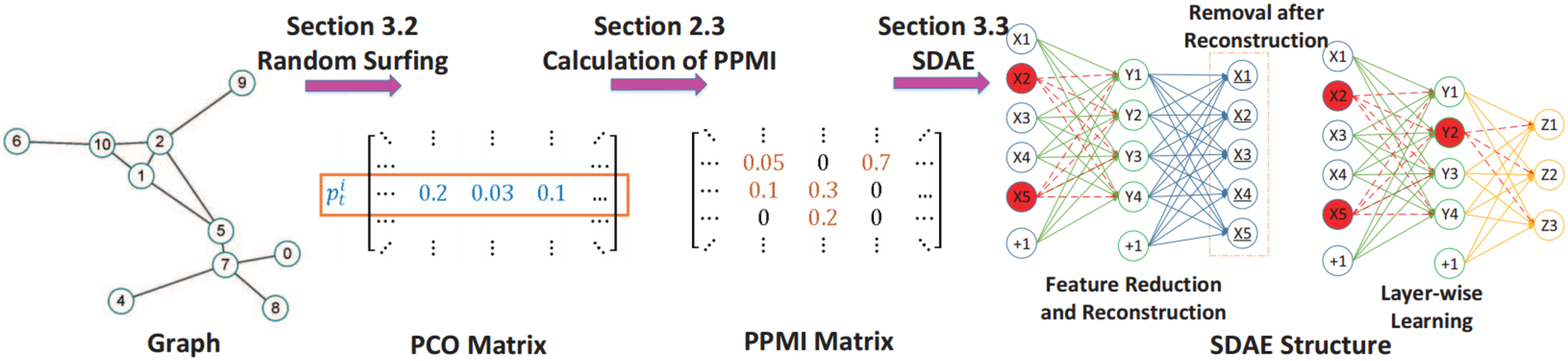}\\
  \caption{Overview of the method proposed by Cao~\textit{et~al.}~\cite{cao2016deep}. Image extracted from~\cite{cao2016deep}. }\label{caoaaai}
\end{figure*}

Cao~\textit{et~al.}~\cite{cao2016deep} propose a network embedding method to capture the weighted graph structure and
represent nodes of non-linear structures.
As shown in Fig.~\ref{caoaaai}, instead of adopting the previous sampling strategy that needs to determine certain hyper parameters, they considers a random surfing model motivated by the
PageRank model. Based on this random surfing model, the representation of a node can be initiatively constructed by combining
the weighted transition probability matrix. After that, the PPMI matrix~\cite{levy2014neural} can be computed. Finally, the stacked denoising autoencoders~\cite{vincent2010stacked} that partially corrupt the input data before taking the training step
are applied to learn the latent representations.

In order to make a general framework on network embedding, Chen~\textit{et~al.}~\cite{chen2017fast} propose a network embedding framework that unifies some of the previous algorithms, such
as LE, DeepWalk and Node2vec. The proposed framework, denoted by GEM-D$[h(\cdot),g(\cdot),d(\cdot ,\cdot)]$, involves three important building blocks:
$h(\cdot)$ is a node proximity function based on the adjacency matrix; $g(\cdot)$ is a warping function that warps the
inner products of network embeddings; and $d(\cdot,\cdot)$ measures the differences between $h$ and $g$. Furthermore,
they demonstrate that the high-order proximity for $h(\cdot)$ and the exponential function for $g(\cdot)$ are more important
for a network embedding algorithm. Based on these observations, they propose UltimateWalk$=$GEM-D$[\prod^{(L)},exp(x),d_{wf}(\cdot,\cdot)]$,
where $\prod^{(L)}$ is a finite-step transition matrix, $exp(x)$ is an exponential function and $d_{wf}(\cdot,\cdot)$ is the warped
Frobenius norm.

In summary, many network embedding methods aim to preserve the local structure of a node, including neighborhood structure, high-order proximity as well as community structure, in the latent low-dimensional space. Both linear and non-linear models are attempted, demonstrating the large potential of deep models in network embedding.

\subsection{Property Preserving Network Embedding}

Among the rich network properties, the properties that are crucial for network inference are the focus in property preserving network embedding.
Specifically, most of the existing property preserving network embedding methods focus on network transitivity in all types of networks and the structural balance property in signed networks.

Ou~\textit{et~al.}~\cite{ou2015non} aim to preserve the non-transitivity property via latent similarity components. The non-transitivity
property declares that, for nodes $A$, $B$ and $C$ in a network where $(A, B)$ and $(B, C)$ are similar pairs, $(A, C)$ may be a dissimilar pair.
For example, in a social network, a student may connect with his classmates and his family, while his
classmates and family are probably very different. To address this, they use a set of linear projection matrices to extract
$M$ hash tables, and thus, each pair of nodes can have $M$ similarities $\{S_{ij}^m\}_{m=1}^M$ based on those hash tables. Then the final similarity between two nodes can be aggregated from $\{S_{ij}^m\}_{m=1}^M$. Finally they
approximate the aggregated similarity to the semantic similarity based on the observation that if two nodes have a large semantic similarity,
at least one of the similarities $S_{ij}^m$ from the hash tables is large, otherwise, all of the similarities are small.

Preserving the asymmetric transitivity property of directed network is considered by
HOPE~\cite{ou2016kdd}. Asymmetric transitivity indicates that, if there is a directed edge from node $i$ to node $j$ and a
directed edge from $j$ to $v$, there is likely a directed edge from $i$ to $v$, but not from $v$ to $i$. In order to
measure this high-order proximity, HOPE summarizes four measurements in a general formulation, that is, Katz Index~\cite{katz1953new},
Rooted PageRank~\cite{liben2007link}, Common Neighbors~\cite{liben2007link}, and Adamic-Adar~\cite{adamic2003friends}.
With the high-order proximity, SVD can be directly
applied to obtain the low dimensional representations. Furthermore, the general formulation of high-order proximity
enables HOPE to
transform the original SVD problem into a generalized SVD problem~\cite{paige1981towards}, such that the time complexity of
HOPE is largely reduced, which means HOPE is scalable for large scale networks.

\begin{figure}
  \centering
  \includegraphics[width=3.20in]{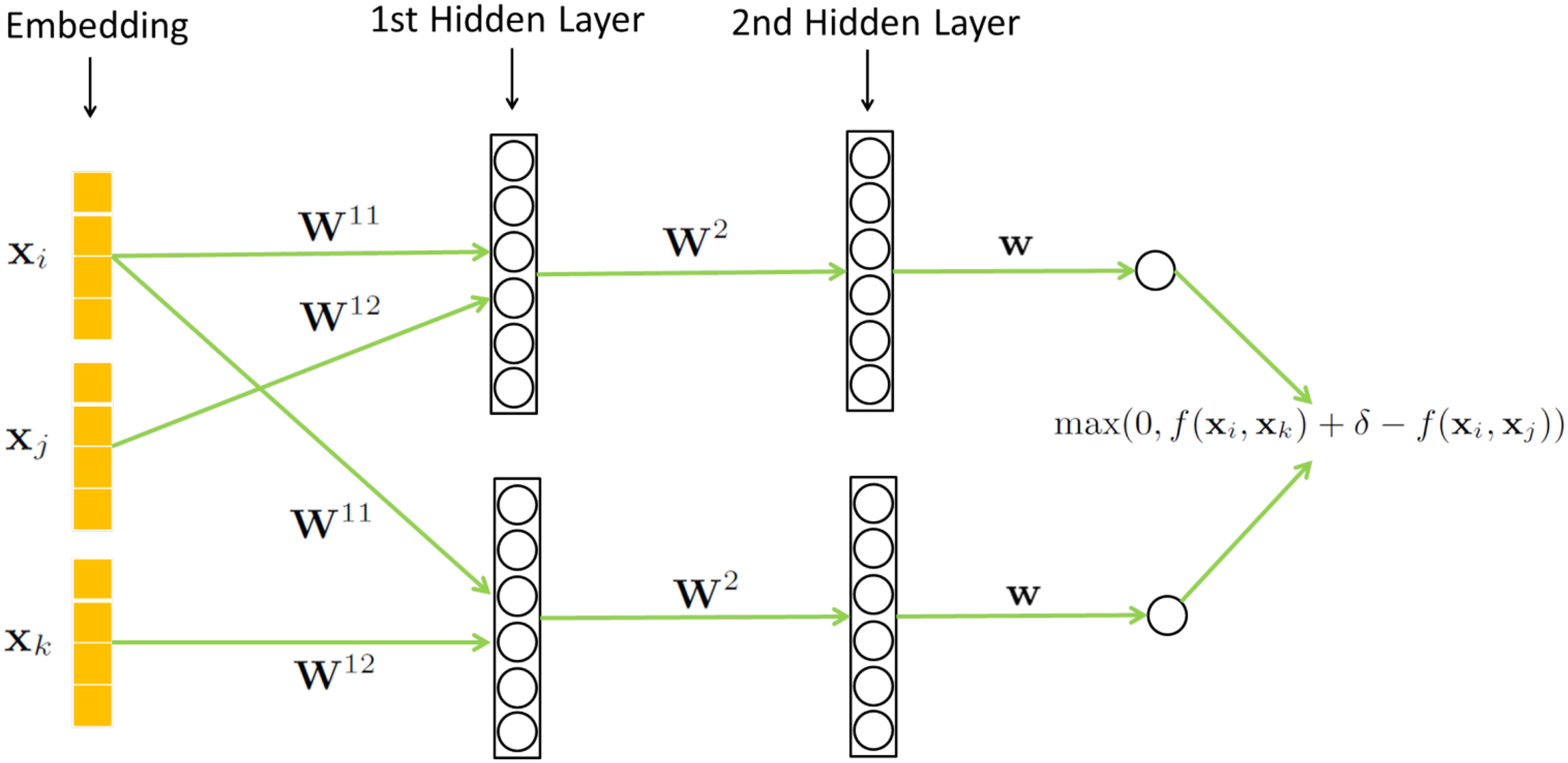}\\
  \caption{The framework of SiNE. Image extracted from~\cite{wang2017signed}. }\label{sidn}
\end{figure}

SiNE~\cite{wang2017signed} is proposed for signed network embedding, which considers both positive and
negative edges in a network. Due to the negative edges, the social theories on signed network, such as structural
balance theory~\cite{cartwright1956structural,cygan2015sitting}, are very different from the unsigned network.
The structural balance theory demonstrates that users in a signed social network should be able to have their
``friends" closer than their ``foes". In other words, given a triplet $(v_i,v_j,v_k)$ with edges $e_{ij}=1$ and $e_{ik}=-1$,
the similarity $f(v_i,v_j)$ between nodes $v_i$ and $v_j$ is larger than $f(v_i,v_k)$. To model the structural balance phenomenon, a deep learning model
consisting of two deep networks
with non-linear functions is designed
to learn the embeddings and preserve the network structure property, which is consistent with the extended structural balance theory. The framework is shown in Fig.~\ref{sidn}.

The methods reviewed in this subsection demonstrate the importance of maintaining network properties in network embedding space, especially the properties that largely affect the evolution and formation of networks. The key challenge in is how to address the disparity and heterogeneity of the original network space and the embedding vector space at property level.

\subsection{Summary}

Generally, most of the structure and property preserving methods take high order proximities of nodes into account, which demonstrate the importance of preserving high order structures in network embedding.
The difference is the strategy of obtaining the high order structures. Some methods implicitly preserve high-order structure by assuming a generative mechanism from a node to its neighbors, while some other methods realize this by explicitly approximating high-order proximities in the embedding space. As topology structures are the most notable characteristic of networks, structure-preserving network methods embody a large part of the literature. Comparatively, property preserving
network embedding is a relatively new research topic and is only studied lightly. As network properties
usually drive the formation and evolution of networks, it shows great potential for future research and applications.

\section{Network Embedding with Side Information}
\label{sec:sideinfo}

Besides network structures, side information is another important information source for network embedding.
Side information in the context of network embedding can be divided into two categories: node content
and types of nodes and edges.
In this section, we review the methods that take side information into network embedding.


\subsection{Network Embedding with Node Content}

In some types of networks, like information networks, nodes are acompanied with rich information, such as node labels, attributes or even semantic descriptions. How to combine them with the network topology in network embedding arouses considerable research interests.

Tu~\textit{et~al.}~\cite{tu2016max} propose a semi-supervised network embedding algorithm, MMDW, by leveraging labeling information of nodes.
MMDW is also based on the DeepWalk-derived matrix factorization. MMDW adopts support vector machines (SVM)~\cite{hearst1998support} and incorporates the label information to find an optimal classifying boundary.
By optimizing the max-margin classifier of SVM and matrix factorization based DeepWalk simultaneously, the representations of nodes that
have more discriminative ability can be learned.

Le~\textit{et~al.}~\cite{le2014probabilistic} propose a generative model for document network embedding, where
the words associated with each documents and the relationships between documents are both considered.
For each node, they learn its low-rank representation $\mathbf{u}_i$ in a low dimensional vector space, which can
reconstruct the network structure. Also, they learn the representation of nodes in the topic space based on the Relational Topic Model (RTM)~\cite{chang2009relational}, where each topic $z$ is associated with a probability distribution over words. To integrate the two aspects, they associate each topic $z$ with a representation $\varphi_z$ in the same low dimensional vector space and
then have the following function:
\begin{equation}\label{lee}
  P(z|v_i)=\frac{exp(-\frac{1}{2}\|\mathbf{u}_i-\varphi_z\|^2)}{\sum_z exp(-\frac{1}{2}\|\mathbf{u}_i-\varphi_z\|^2)}.
\end{equation}
Finally, in a unified generative process, the representations of nodes $\mathbf{U}$ can be learned.


Besides network structures, Yang~\textit{et~al.}~\cite{yang2015network} propose TADW that takes the rich information (e.g., text) associated with nodes into account when they learn
the low dimensional representations of nodes. They first prove that DeepWalk is equivalent to factorizing the matrix $\mathbf{M}$ whose
element $M_{ij}=\log ([e_i(\mathbf{A}+\mathbf{A}^2+...+\mathbf{A}^t)]_j/t)$, where $\mathbf{A}$ is the adjacency matrix, $t$ denotes the $t$
steps in a random walk and $e_i$ is a row vector where all entries are 0 except the $i$-th entry is 1.
Then, based on the DeepWalk-derived matrix factorization and motivated by the inductive matrix completion~\cite{natarajan2014inductive}, they incorporate rich text information $\mathbf{T}$ into network embedding as follows:
\begin{equation}\label{tadw}
  \min_{\mathbf{W,H}}\|\mathbf{M-W}^T\mathbf{HT}\|_F^2+\frac{\lambda}{2}(\|\mathbf{W}\|_F^2+\|\mathbf{H}\|_F^2).
\end{equation}
Finally, they concatenate the optimal $\mathbf{W}$ and $\mathbf{HT}$ as the representations of nodes.

\begin{figure}
  \centering
  \includegraphics[width=3.20in]{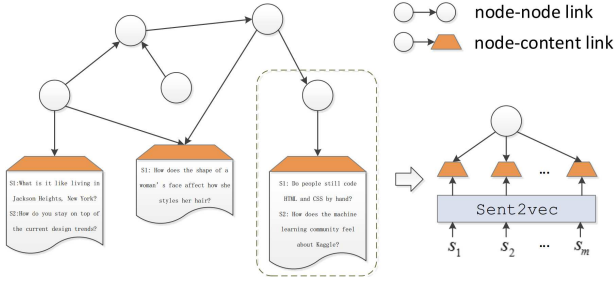}\\
  \caption{The augmented network proposed by Sun~\textit{et~al.}~\cite{sun2016general}. Image extracted from~\cite{sun2016general}. }\label{agf}
\end{figure}

TADW suffers from high computational cost and the node attributes just simply incorporated as unordered features lose the much semantic information. Sun~\textit{et~al.}~\cite{sun2016general} consider the content as a special kind of nodes, and give rise to an augmented network, as shown in Fig.~\ref{agf}. With this augmented network,
they are able to model the node-node links and node-content links in the latent vector space. They use a logistic function
to model the relationship in the new augmented network, and by combining with negative sampling, they can learn the representations of nodes
in a joint objective function, such that the representations can preserve the network structure as well as the relationship between
the node and content.

\begin{figure}
  \centering
  \includegraphics[width=2.60in]{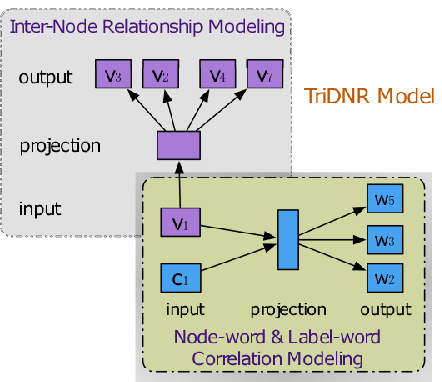}\\
  \caption{The framework of TriDNR~\cite{pan2016tri}. Image extracted from~\cite{pan2016tri}. }\label{tridnr}
\end{figure}

Pan~\textit{et~al.}~\cite{pan2016tri} propose a coupled deep model that incorporates network structure, node attributes and node labels
into network embedding. The architecture of the proposed model is shown in Fig.~\ref{tridnr}. Consider a network with $N$ nodes $\{v_i\}_{i=1,...,N}$, where each node is associated with a set of
words $\{w_i\}$, and some nodes may have $|L|$ labels $\{c_i\}$. To exploit this information, they aim to maximize the following function:
\begin{equation}\label{eq:tridnr}
\begin{split}
   \mathcal{L}=& (1-\alpha)\sum_{i=1}^N\sum_{s\in \mathcal{S}}\sum_{-b\leq j\leq b,j\neq 0}\log P(v_{i+j}|v_i) \\
    & \alpha \sum_{i=1}^N\sum_{-b\leq j\leq b}\log P(w_j|v_i)+\alpha \sum_{i=1}^{|L|}\sum_{-b\leq j\leq b}\log P(w_j|c_i),
\end{split}
\end{equation}
where $\mathcal{S}$ is the random walks generated in the network and $b$ is the window size of sequence. Specifically, function $P$,
which captures the probability of observing contextual nodes (or words) given the current node (or label),
can be computed using the soft-max function. In Eq.~\ref{eq:tridnr}, the first term is also motivated by Skip-Gram, similar to DeepWalk,
which models the network structure.
The second term models the node-content correlations and the third term models the label-node correspondences. As a result,
the learned representations is enhanced by network structure, node content, and node labels.

\begin{figure*}
  \centering
  \includegraphics[width=6.30in]{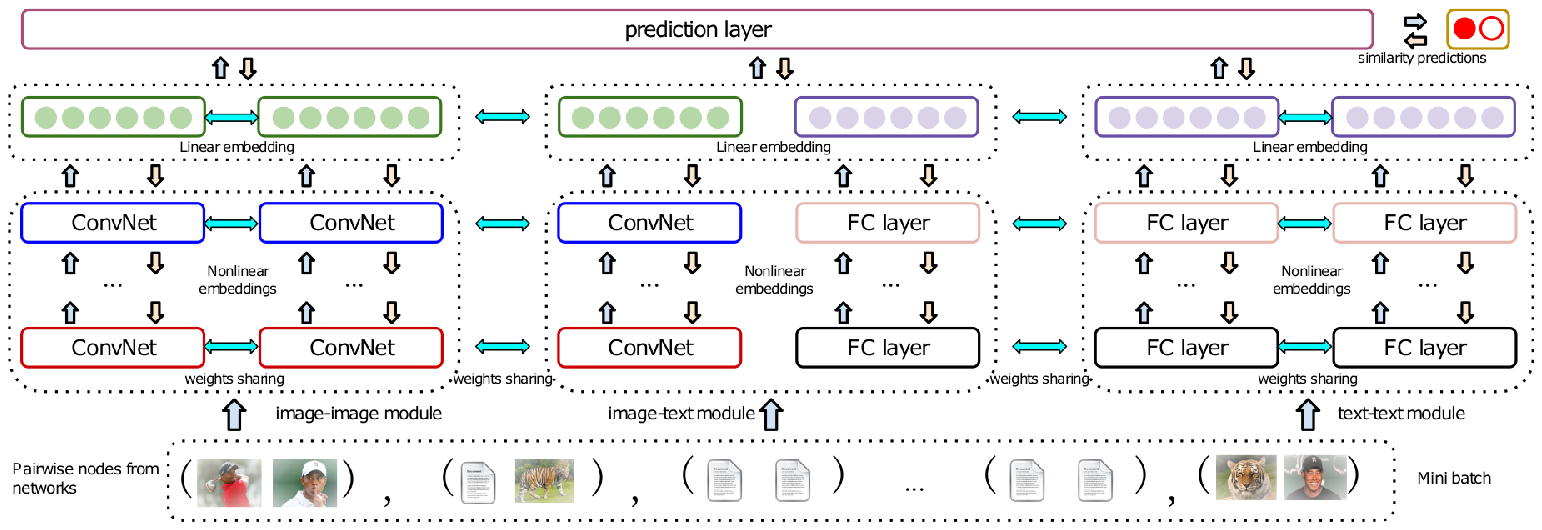}\\
  \caption{Overview of the method proposed by Chang~\textit{et~al.}~\cite{chang2015heterogeneous}. Image extracted from~\cite{chang2015heterogeneous}. }\label{hin}
\end{figure*}

LANE~\cite{huang2017label} is also proposed to incorporate the label information into the attributed network embedding.
Unlike the previous network embedding methods, LANE is mainly based on spectral techniques~\cite{chung1997spectral}. LANE adopts the cosine similarity to
construct the corresponding affinity matrices of the node attributes, network structure, and labels. Then, based on the
corresponding Laplacian matrices, LANE is able to
map the three different sources into different latent representations, respectively. In order to build the relationship
among those three representations, LANE projects all these latent representations into a new common space by leveraging the variance of the
projected matrix as the correlation metric. The learned representations of nodes are able to capture
the structure proximities as well as the correlations in the label informed attributed network.

Although different methods adopt different
strategies to integrate node content and network topology,
they all assume that node content provides additional
proximity information to constrain the representations of nodes.

\subsection{Heterogeneous Information Network Embedding}

Different from networks with node content, heterogeneous networks consist of different types of nodes and links. How to unify the heterogeneous types of nodes and links in network embedding is also an interesting and challenging problem.

Yann~\textit{et~al.}~\cite{jacob2014learning} propose a heterogeneous social network embedding algorithm for classifying nodes.
They learn the representations of all types of nodes in a common vector space, and perform the inference in
this space.
In particular, for the node $\mathbf{u}_i$ with type $t_i$, they utilize a linear classification function $f_{\theta}^{t_i}$ to predict its label and adopt
the hinge-loss function $\Delta$ to measure the loss with the true label $y_i$:
\begin{equation}\label{nc1}
  \sum_{i=1}^l\Delta (f_{\theta}^{t_i}(\mathbf{u}_i),y_i),
\end{equation}
where $l$ is the number of labeled nodes. To preserve the local structures in the latent space,
they impose the following smoothness constraint, which enforces that two nodes $i$ and $j$ will be
close in the latent space if they have a large weight $W_{ij}$ in the heterogeneous network:
\begin{equation}\label{nc2}
  \sum_{i,j}W_{ij}\|\mathbf{u}_i-\mathbf{u}_j\|^2.
\end{equation}
In this way, different types of nodes are mapped into a common latent space. The overall loss function combines the classification and regularization
losses Eq.~\eqref{nc1} and Eq.~\eqref{nc2}. A stochastic gradient descent method is used here to learn the representations of nodes in a heterogeneous network for classifying.

Chang~\textit{et~al.}~\cite{chang2015heterogeneous} propose a deep embedding algorithm for heterogeneous networks, whose nodes
have various types. The main goal of the heterogeneous network embedding is to learn the representations of nodes with different types
such that the heterogeneous network structure can be well preserved.  As shown in Fig.~\ref{hin}, given a heterogeneous network
with two types of data (e.g., images and texts), there are three types of edges, i.e., image-image, text-text, and image-text.
The nonlinear embeddings of images and texts are learned by a CNN model and the fully connected layers, respectively.
By cascading the extra linear embedding layer, the representations of images and texts can be mapped to
a common space. In the common space, the similarities between data from different modalities can be directly measured, so that
if there is an edge in the original heterogeneous network, the pair of data has similar representations.

Huang and Mamoulis~\cite{huang2017heterogeneous} propose a meta path similarity preserving heterogeneous information network embedding algorithm.
To model a particular relationship, a meta path~\cite{sun2011pathsim} is a sequence of object types with edge types in between. They develop a fast dynamic programming approach to calculate the truncated meta path based proximities, whose time
complexity is linear to the size of the network. They adopt a similar strategy as LINE~\cite{tang2015line} to preserve
the proximity in the low dimensional space.

Xu~\textit{et~al.}~\cite{xu2017embedding} propose a network embedding method for coupled heterogeneous network. The coupled
heterogeneous network consists of two different but related homogeneous networks. For each homogeneous network, they
adopt the same function (Eq.~\eqref{line1}) as LINE to model the relationships between nodes. Then the harmonious embedding matrix is introduced to measure the
closeness between nodes of different networks. Because the inter-network edges are able to provide the complementary information in
the presence of intra-network edges, the learned embeddings of nodes also perform well on several tasks.

\subsection{Summary}

In the methods preserving side information, side information introduces additional proximity measures so that the relationships between nodes can be learned more comprehensively. Their difference is the way of integrating network structures
and side information. Many of them are naturally extensions from structure preserving network embedding methods.

\section{Advanced Information Preserving Network Embedding}
\label{sec:advanced}

In this section, we review network embedding methods that take additional advanced information into account so as to solve
some specific analytic tasks. Different from side information, the advanced information refers to the supervised or pseudo supervised information in a specific task.

\subsection{Information Diffusion}

Information diffusion~\cite{guille2013information} is an ubiquitous phenomenon on the web, especially in social networks. Many real applications, such as marketing, public opinion formation, epidemics, are related to information diffusion. Most of the previous studies on information diffusion are conducted in original network spaces.

Recently, Simon~\textit{et~al.}~\cite{bourigault2014learning} propose a social network embedding algorithm for predicting information
diffusion. The basic idea is to map the observed information diffusion process into a heat diffusion process modeled
by a diffusion kernel in
the continuous space. Specifically, the diffusion kernel in a $d$-dimensional Euclidean space is defined as
\begin{equation}\label{id1}
  K(t,j,i)=(4\Pi t)^{-\frac{d}{2}}e^{-\frac{\|j-i\|^2}{4t}}.
\end{equation}
It models the heat at location $i$ at time $t$ when an initial unit heat is positioned at location $j$, which also
models how information spreads between nodes in a network.

The goal of the proposed algorithm is to learn the representations of nodes in the latent space such that
the diffusion kernel can best explain the cascades in the training set. Given the representation $\mathbf{u}_j$ of
the initial contaminated node $j$ in cascade $c$, the contamination score of node $i$ can be computed by
\begin{equation}\label{id2}
  K(t,j,i)=(4\Pi t)^{-\frac{d}{2}}e^{-\frac{\|\mathbf{u}_j-\mathbf{u}_i\|^2}{4t}}.
\end{equation}
The intuition of Eq.~\eqref{id2} is that the closer a node in the latent space is from the source node,
the sooner it is infected by information from the source node. As the cascade $c$ offers a guidance for the information diffusion of nodes, we expect the contamination score to
be as closely consistent with $c$ as possible, which gives rise to the following empirical risk function:
\begin{equation}\label{id3}
  L(\mathbf{U})=\sum_c \Delta (K(.,j,.),c),
\end{equation}
where function $\Delta$ is a measure of the difference between the predicted score and the observed diffusion in $c$.
By minimizing the Eq.~\eqref{id3} and reformulating it as a ranking problem, the optimal representations $\mathbf{U}$ of
nodes can be obtained.


\begin{figure*}
  \centering
  \includegraphics[width=6.20in]{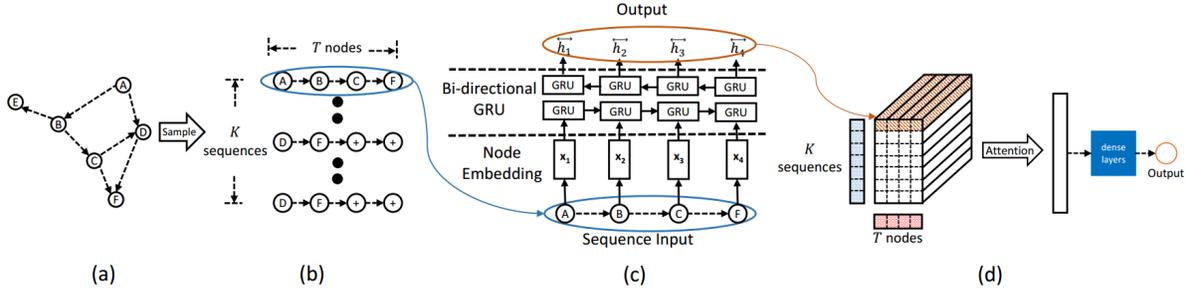}\\
  \caption{The end-to-end pipeline of DeepCas proposed by Li~\textit{et~al.}~\cite{li2017deepcas}. Image extracted from~\cite{li2017deepcas}. }\label{deepcas}
\end{figure*}

The cascade prediction problem here is defined as predicting the increment of cascade size after a given
time interval~\cite{li2017deepcas}. Li~\textit{et~al.}~\cite{li2017deepcas} argue that the previous work on cascade prediction all depends on
the bag of hand-crafting features to represent the cascade and network structures. Instead, they present an end-to-end
deep learning model to solve this
problem using the idea of network embedding, as illustrated in Fig.~\ref{deepcas}. Similar to DeepWalk~\cite{perozzi2014deepwalk}, they perform a random walk over a cascade graph to sample
a set of paths. Then the Gated Recurrent Unite (GRU)~\cite{hochreiter1997long}, a specific type of recurrent neural network~\cite{mikolov2010recurrent}, is applied to these
paths and learn the embeddings for these paths. The attention mechanism is then used to assemble these embeddings
to learn the representation of this cascade graph. Once the representation of this cascade is known, a
multi-layer perceptron~\cite{ruck1990multilayer} can be adopted to output the final predicted size of this cascade.
The whole procedure is able to learn the representation of cascade graph in an end-to-end manner. The experimental results
on the Twitter and Aminer networks show promising performance on this task.

\subsection{Anomaly Detection}

\begin{figure}
  \centering
  \includegraphics[width=3.20in]{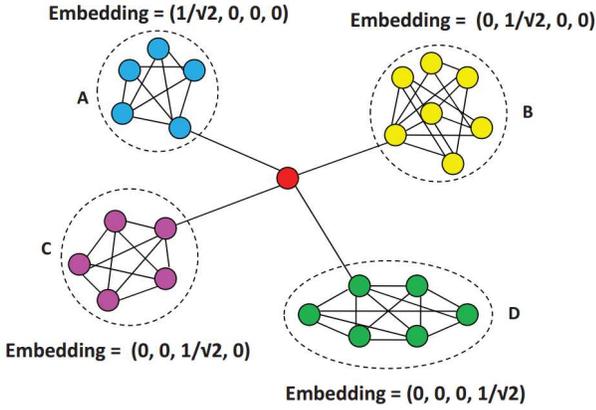}\\
  \caption{The anomalous (red) nodes in embedding, and A, B, C, D are four communities~\cite{hu2016embedding}. Image extracted from~\cite{hu2016embedding}. }\label{anomaly}
\end{figure}


Anomaly detection has been widely investigated in previous work~\cite{akoglu2015graph}. Anomaly detection in networks
aims to infer the structural inconsistencies, which means the anomalous nodes that connect to various diverse influential
communities~\cite{burt2004structural,hu2016embedding}, such as the red node in Fig.~\ref{anomaly}. Hu~\textit{et~al.}~\cite{hu2016embedding} propose a network embedding
 based method for anomaly detection. In particular, in the proposed model, the $k$-th element $u_i^k$ in the embedding $\mathbf{u}_i$ of
node $i$ represents the correlation between node $i$ and community $k$. Then, they assume that the community memberships of two linked nodes should be similar. Therefore, they can minimize the following
objective function:
\begin{equation}\label{ad}
  L=\sum_{(i,j)\in E}\|\mathbf{u}_i-\mathbf{u}_j\|^2+\alpha \sum_{(i,j)\notin E}(\|\mathbf{u}_i-\mathbf{u}_j\|-1)^2.
\end{equation}
This optimization problem can be solved by the gradient descent method. By taking the neighbors of a node into
account, the embedding of the node can be obtained by a weighted sum of the embeddings of all its neighbors. An anomaly node in this context is one connecting to a set of different communities. Since the learned embedding of nodes
captures the correlations between nodes and communities, based on the embedding, they propose a new measure to indicate
the anomalousness level of a node. The larger the value of the measure, the higher the propensity for a node being an anomaly node.

\subsection{Network Alignment}

\begin{figure}
  \centering
  \includegraphics[width=3.00in]{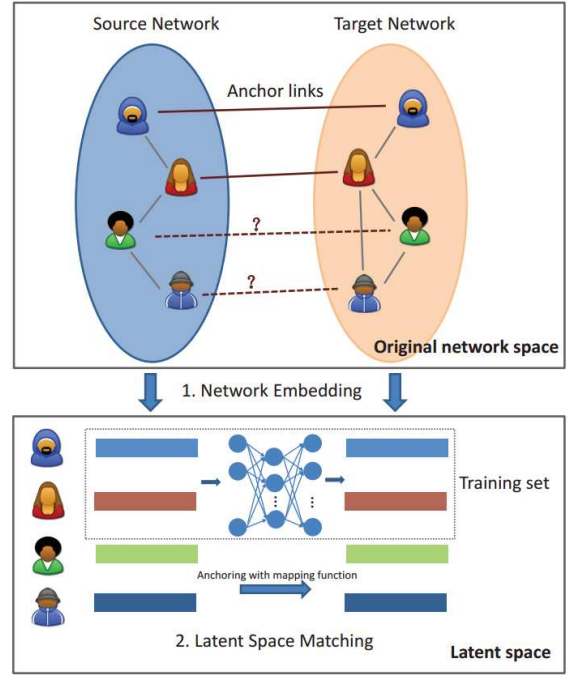}\\
  \caption{The illustrative diagram of network embedding for anchor link prediction proposed by Man~\textit{et~al.}~\cite{man2016predict}. Image extracted from~\cite{man2016predict}. }\label{anchor}
\end{figure}

The goal of network alignment is to establish the correspondence between the nodes from two networks. Man~\textit{et~al.}~\cite{man2016predict} propose a network embedding algorithm to predict the anchor links across social networks.
The same users who are shared by different social networks naturally form the anchor links, and these links bridge the different
networks. As illustrated in Fig.~\ref{anchor}, the anchor link prediction problem is, given source network $G^s$ and target network $G^t$ and a set of
observed anchor links $T$, to identify the hidden anchor links across $G^s$ and $G^t$.

First, Man~\textit{et~al.}~\cite{man2016predict} extend the original sparse networks
$G^s$ and $G^t$ to the denser networks. The basic idea is that given a pair of users with anchor links, if they have a connection in one network, so
do their counterparts in the other network~\cite{bayati2009algorithms}, in this way, more links will be added to the original networks. For a pair of nodes $i$ and $j$ whose representations
are $\mathbf{u}_i$ and $\mathbf{u}_j$, respectively, by combining the negative sampling strategy, they
use the following function to preserve the structures of $G^s$ and $G^t$ in a vector space:
\begin{equation}\label{alp1}
  \log \sigma(\mathbf{u}_i^T\mathbf{u}_j)+\sum_{k=1}^KE_{v_k\propto P_n(v)}[\log (1-\sigma(\mathbf{u}_i^T\mathbf{u}_k))],
\end{equation}
where $\sigma (x)=1/(1+exp(-x))$. The first term models the observed edges, and the second term samples $K$ negative edges.

Then given the observed anchor links $(v_i^s,u_j^t)\in T$ and the representations $\mathbf{u}_i$ and $\mathbf{u}_j$, they
aim to learn a mapping function $\phi$ parameterized by $\theta$ so as to bridge these two representations. The loss function
is defined as:
\begin{equation}\label{alp2}
  \|\phi (\mathbf{u}_i;\theta)-\mathbf{u}_j\|_F.
\end{equation}
The mapping function can be linear or non-linear via Multi-Layer Perceptron (MLP)~\cite{ruck1990multilayer}. By optimizing
Eq.~\eqref{alp1} and Eq.~\eqref{alp2} simultaneously, the representations that can preserve the network structure and respect the
observed anchor links can be learned.

\subsection{Summary}

Advanced information preserving network embedding usually consists of two parts. One is to preserve the network structure so as to learn the representations of nodes. The other is to establish the connection between the representations of nodes and the target task.
The first one is similar to structure and property preserving network embedding, while the second one usually
needs to consider the domain knowledge of a specific task. The domain knowledge encoded by the advanced information makes it possible to develop end-to-end solutions for network applications. Compared with the hand-crafted network features, such as numerous network centrality measures, the combination of advanced information and network embedding techniques enables representation learning for networks. Many network applications may be benefitted from this new paradigm.


\section{Network Embedding in Practice}
\label{sec:app}

In this section, we summarize the data sets, benchmarks, and evaluation tasks that are commonly used in developing new network embedding methods.

\begin{table*}
\caption{A summary of real world networks}\label{asrw}
\newcommand{\tabincell}[2]{\begin{tabular}{@{}#1@{}}#2\end{tabular}}
\centering
\begin{tabular}{|c|c|c|c|c|c|c|}
\hline
networks & \tabincell{c}{structure and property \\preserving network\\ embedding} & \tabincell{c}{network embedding\\ with side\\ information} & classification & link prediction & clustering & visualization \\
\hline
 BLOGCATALOG & $\surd$ & & $\surd$ & $\surd$ & $\surd$ & $\surd$ \\
 \hline
 FLICKR & $\surd$ &  & $\surd$ & $\surd$ & $\surd$ & $\surd$ \\
 \hline
 YOUTUBE & $\surd$ & & $\surd$ & $\surd$ & $\surd$ & $\surd$ \\
 \hline
 Twitter & $\surd$ & & & $\surd$ & &  \\
 \hline
 DBLP &$\surd$  & $\surd$ & $\surd$ & $\surd$ & $\surd$ & $\surd$ \\
 \hline
 Cora & $\surd$ & $\surd$ & $\surd$ & $\surd$ & $\surd$ & $\surd$ \\
 \hline
 Citeseer & $\surd$ & $\surd$ & $\surd$ & $\surd$ & $\surd$ & $\surd$ \\
 \hline
 ArXiv & $\surd$ & & & $\surd$ & &  \\
 \hline
 Wikipedia & $\surd$ & & $\surd$ & $\surd$ & $\surd$ & $\surd$ \\
 \hline
 PPI & $\surd$ & & & $\surd$ & &  \\
\hline
\end{tabular}
\end{table*}

\subsection{Real World Data Sets}
Getting real network data sets in academic research is always far from trivial. Here, we describe some most popular real world networks currently used in
network embedding literature. The data sets can be roughly divided into four groups according to the nature of the networks:
social networks, citation networks, language networks, and biological networks. A summary of these data sets can be found in
Table~\ref{asrw}. Please note that,
the same name may be used to refer to different variants in different studies. Here we aim to provide an overview of the networks, and do not
attempt to describe all of those variants in detail.

\subsubsection{Social Networks}

\begin{itemize}
\item \textbf{BLOGCATALOG}~\cite{tang2009relational}. This is a network of social relationships of the bloggers listed on the BlogCatalog website.
One instance of this data set can be found at \url{http://socialcomputing.asu.edu/datasets/BlogCatalog3}.

\item \textbf{FLICKR}~\cite{tang2009relational}. This is a network of the contacts between users of the photo sharing websites Flickr.
One instance of the network can be downloaded at \url{http://socialcomputing.asu.edu/datasets/Flickr}.

\item \textbf{YOUTUBE}~\cite{tang2009scalable}. This is a network between users of the popular video sharing website, Youtube.
One instance of the network can be found at \url{http://socialcomputing.asu.edu/datasets/YouTube2}.

\item \textbf{Twitter}~\cite{de2010does}. This is a network between users on a social news website Twitter.
One instance of the network can be downloaded at \url{http://socialcomputing.asu.edu/datasets/Twitter}.
\end{itemize}

\subsubsection{Citation Networks}
\begin{itemize}
\item \textbf{DBLP}~\cite{tang2008arnetminer}. This network represents the citation relationships between
authors and papers. One instance of the data set can be found at \url{http://arnetminer.org/citation}.

\item \textbf{Cora}~\cite{mccallum2000automating}. This network represents the citation relationships between
scientific publications. Besides the link information, each publication is also associated with a word vector indicating the
absence/presence of the corresponding words from the dictionary. One instance of the data set can be found at \url{https://linqs.soe.ucsc.edu/node/236}.

\item \textbf{Citeseer}~\cite{mccallum2000automating}. This network, similar to Cora, also consists of scientific publications and their citation
relationships. One instance of the data set can be downloaded at \url{https://linqs.soe.ucsc.edu/node/236}.

\item \textbf{ArXiv}~\cite{leskovec2007graph,leskovec2016snap}. This is the collaboration network constructed from the ArXiv website.
One instance of the data set can be found at \url{http://snap.stanford.edu/data/ca-AstroPh.html}.
\end{itemize}

\subsubsection{Language Networks}
\begin{itemize}
\item \textbf{Wikipedia}~\cite{mahoney2011large}. This is a word co-occurrence network from the English Wikipedia pages. One instance of the data set can be
found at \url{http://www.mattmahoney.net/dc/textdata}.
\end{itemize}
\subsubsection{Biological Networks}
\begin{itemize}
\item \textbf{PPI}~\cite{breitkreutz2007biogrid}. This is a subgraph of the biological network that represents the pairwise
physical interactions between proteins in yeast. One instance of the data set can be downloaded at \url{http://konect.uni-koblenz.de/networks/maayan-vidal}.
\end{itemize}

\subsection{Node Classification}

Given some nodes with known labels in a network, the node classification problem is to classify the rest nodes into different classes.
Node classification is one of most primary applications for network embedding~\cite{perozzi2014deepwalk,tang2015line}.
Essentially, node classification based on network embedding for can be divided into three steps. First, a network embedding algorithm is applied to embed the network into a low dimensional space. Then, the nodes with known labels are used as the training set. Last, a classifier, such as Liblinear~\cite{fan2008liblinear}, is learned from the training set. Using the trained classifier, we can infer the labels of the rest nodes.

The popularly used evaluation metrics for multi-label classification problem include Micro-F1 and Macro-F1~\cite{tang2009relational}. Specifically, for an overall label set $\mathcal{C}$ and a label $A$, let $TP(A)$, $FP(A)$, and $FN(A)$ be the number of true positives, false positives, and false negatives in the instances predicted as $A$, respectively. Then the Micro-F1 is defined as
 \begin{equation}\label{ncc1}
 \begin{split}
      &Pr=\frac{\sum_{A\in \mathcal{C}}TP(A)}{\sum_{A\in \mathcal{C}}(TP(A)+FP(A))},\\
      &R=\frac{\sum_{A\in \mathcal{C}}TP(a)}{\sum_{A\in \mathcal{C}}(TP(A)+FN(A))}, \\
     & \text{Micro-F1}=\frac{2*Pr*R}{Pr+R}.
 \end{split}
 \end{equation}
 The Macro-F1 measure is defined as
 \begin{equation}\label{ncc2}
   \text{Macro-F1}=\frac{\sum_{A\in \mathcal{C}}F1(A)}{|\mathcal{C}|},
 \end{equation}
 where $F1(A)$ is the F1-measure for the label $A$.

The multi-label classification application has been successfully tested on four categories of data sets, namely social networks (BLOGCATALOG~\cite{tang2009relational}, FLICKR~\cite{tang2009relational}, and YOUTUBE~\cite{tang2009scalable}), citation networks
 (DBLP~\cite{tang2008arnetminer}, Cora~\cite{mccallum2000automating}, and Citeseer~\cite{mccallum2000automating}),
 language networks (Wikipedia~\cite{mahoney2011large}), and biological networks (PPI~\cite{breitkreutz2007biogrid}).

Specifically, a social network usually is a communication network among users on online platforms. DeepWalk~\cite{perozzi2014deepwalk}, GraRep~\cite{cao2015grarep}, SDNE~\cite{wang2016kdd}, node2vec~\cite{jure2016kdd}, and LANE~\cite{huang2017label} conduct classification on BLOGCATALOG to evaluate the performance. Also, the classification performance on FLICKR has been assessed in~\cite{perozzi2014deepwalk,tang2015line,wang2016kdd,huang2017label}.
Some studies~\cite{perozzi2014deepwalk,tang2015line,wang2016kdd} apply their algorithms to the Youtube network, which also achieves promising classification results.
A citation network usually represents the citation relationships between authors or between papers. For example,
~\cite{tang2015line,pan2016tri} use the DBLP network to test the classification performance. Cora is used in
~\cite{yang2015network,tu2016max}. Citeseer is used in~\cite{yang2015network,pan2016tri,tu2016max}. The classification performance on language networks, such as Wikipedia, is also widely studied~\cite{tang2015line,jure2016kdd,yang2015network,tu2016max}.
The Protein-Protein Interactions (PPI) is used in~\cite{jure2016kdd}. Based on NUS-WIDE~\cite{chua2009nus}, a heterogeneous network extracted from Flickr, Chang~\textit{et~al.}~\cite{chang2015heterogeneous}
 validated the superior classification performance of network embedding on heterogeneous networks.

 To summarize, network embedding algorithms have been widely used on various networks and have been well demonstrated their effectiveness on node classification.

 \begin{figure*}
  \centering
  \includegraphics[width=7.00in]{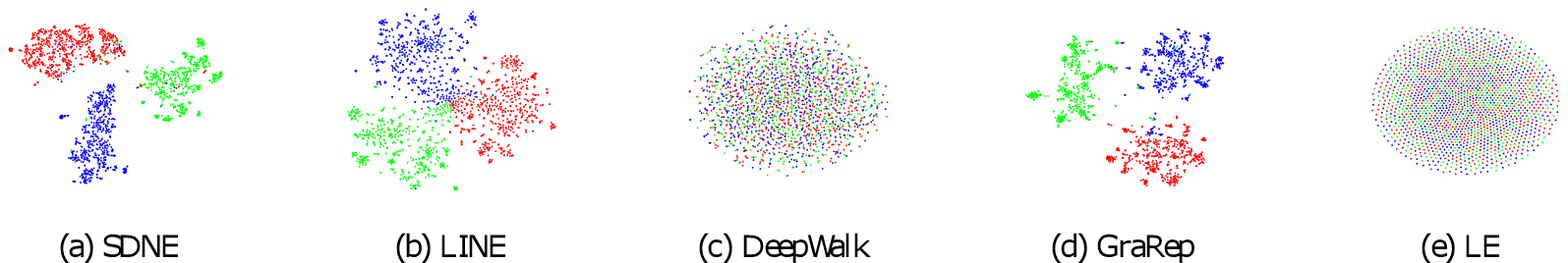}\\
  \caption{Network visualization of 20-NewsGroup by different network embedding algorithms, i.e., SDNE~\cite{wang2016kdd}, LINE~\cite{tang2015line}, DeepWalk~\cite{perozzi2014deepwalk}, GraRep~\cite{cao2015grarep}, LE~\cite{belkin2003laplacian}. Image extracted from SDNE~\cite{wang2016kdd}. }\label{nv}
\end{figure*}

\subsection{Link Prediction}

Link prediction, as one of the most fundamental problems on network analysis, has received a considerable amount of attention~\cite{liben2007link,lu2011link}. It aims to estimate the likelihood of the existence of an edge between two nodes based on observed network structure~\cite{getoor2005link}.
Since network embedding algorithms are able to learn the vector based features for each node, the similarity between nodes can be easily
estimated, for example, by the inner product or the cosine similarity. A larger similarity implies that the two nodes may have a higher propensity to be linked.

Generally, precision$@k$ and Mean Average Precision (MAP) are used to evaluate the link prediction performance~\cite{wang2016kdd}, which are defined as follows.
\begin{equation}\label{pk}
  precision@k(i)=\frac{|\{j|i,j\in V, index(j)\leq k, \triangle_i(j)=1\}|}{k},
\end{equation}
where $V$ is the set of nodes, $index(j)$ is the ranked index of the $j$-th node and $\triangle_i(j)=1$ indicates that nodes $i$ and $j$
have an edge.
\begin{equation}\label{map}
\begin{split}
  &AP(i)=\frac{\sum_jprecision@j(i)*\triangle_i(j)}{|\{\triangle_i(j)=1\}|},\\
  &MAP=\frac{\sum_i\in QAP(i)}{|Q|},
\end{split}
\end{equation}
where $Q$ is the query set.

The popularly used real networks for the link prediction task can be divided into three categories: citation networks (ARXIV~\cite{leskovec2007graph,leskovec2016snap}
and DBLP\footnote{\url{http://dblp.uni-trier.de/}}), social networks (SN-TWeibo\footnote{\url{http://www.kddcup2012.org/c/kddcup2012-track1/data}}, SN-Twitter~\cite{de2010does},
Facebook~\cite{leskovec2016snap}, Epinions\footnote{\url{http://www.epinions.com/}}, and Slashdot\footnote{\url{http://slashdot.org/}}), and biological networks
(PPI~\cite{breitkreutz2007biogrid}). Specifically,
\cite{wang2016kdd} and~\cite{jure2016kdd} test the effectiveness on ARXIV\footnote{\url{https://arxiv.org/}}. HOPE~\cite{ou2016kdd} applies network embedding to link prediction on two directed networks SN-Twitter, which is a subnetwork of Twitter\footnote{https://twitter.com/}, and SN-TWeibo, which is a subnetwork of the social network in Tencent Weibo\footnote{\url{http://t.qq.com/}}. Node2vec~\cite{jure2016kdd} tests the performance of link prediction on a social network Facebook and a biological network PPI. EOE~\cite{xu2017embedding} uses DBLP to demonstrate the effectiveness on citation networks.
Based on two social networks, Epinions and Slashdot, SiNE~\cite{wang2017signed} shows the superior performance of signed network
embedding on link prediction.

To sum up, network embedding is able to capture inherent network structures, and thus naturally it is suitable for
link prediction applications. Extensive experiments on various networks have demonstrated that network embedding can tackle link prediction effectively.

\subsection{Node Clustering}

Node clustering is to divide the nodes in a network into clusters such that the nodes within the same cluster are more similar to each other than the nodes in different clusters.
Network embedding algorithms learn representations of nodes in low dimensional vector spaces, so many typical clustering methods, such as Kmeans~\cite{macqueen1967some}, can be directly
adopted to cluster nodes based on their learned representations.

Many evaluation criteria have
been proposed for clustering evaluation. Accuracy (AC) and normalized mutual information (NMI)~\cite{cai2011graph} are frequently used to assess
the clustering performance on graphs and networks. Specifically, AC is used to measure the percentage of correct labels obtained.
Given $n$ data, let $l_i$ and $r_i$ be the obtained cluster label and the ground truth label, respectively.
AC is defined as
\begin{equation}\label{ac}
  AC=\frac{\sum_{i=1}^n\delta (r_i,map(l_i))}{n},
\end{equation}
where $\delta (x,y)$ equals one if $x=y$ and equals zero otherwise, and
$map(l_i)$ is the permutation mapping function that maps each cluster label $l_i$ to
the equivalent label from the data, which can be found using the Kuhn-Munkres algorithm~\cite{lovasz2009matching}.

Given the set of clusters obtained from the ground truth and obtained from the algorithm, respectively, denoted
by $C$ and $C'$, the NMI can be defined as
\begin{equation}\label{nmi}
  NMI(C,C')=\frac{MI(C,C')}{max(H(C),H(C'))},
\end{equation}
where $H(C)$ is the entropy of $C$, and $MI(C,C')$ is the mutual information metric of $C$ and $C'$.

The node clustering performance is tested on three types of networks: social networks (e.g., Facebook~\cite{traud2012social} and YELP~\cite{huang2017heterogeneous}), citation networks (e.g., DBLP~\cite{sun2011pathsim}), and document networks (e.g., 20-NewsGroup~\cite{tian2014learning}).
In particular,~\cite{chang2015heterogeneous} extracts a social network from a social blogging site. It uses
the TF-IDF features extracted from the blogs as the features of blog users and the ``following" behaviors to construct the linkages. It successfully applies network embedding to the node clustering task.
\cite{wang2017community} uses the Facebook social network to demonstrate the effectiveness of community preserving
network embedding on node clustering. \cite{huang2017heterogeneous} is applied to more social networks including MOVIE, a network extracted
from YAGO~\cite{huang2016meta} that contains knowledge about movies, YELP, a network extracted from YELP that is about reviews given to restaurants, and GAME, extracted from
Freebase~\cite{bollacker2008freebase} that is related to video games.
\cite{cao2016deep} tests the node clustering performance on a document network, 20-NewsGroup network, which consists of documents.
The node clustering performance on citation networks is tested~\cite{huang2017heterogeneous} by clustering authors in DBLP. The results show the superior clustering performance on citation networks.

In summary, node clustering based on network
embedding is tested on different types of networks. Network embedding has become an effective method to solve the node clustering problem.

\subsection{Network Visualization}

Another important application of network embedding is network visualization, that is, generating meaningful visualization that layouts a network on a two dimensional space.
By applying the visualization tool, such as t-SNE~\cite{maaten2008visualizing}, to the learned low dimensional
representations of nodes, it is easy for users to see a big picture of a sophisticated network so that the community structure or node centrality
can be easily revealed.

More often than not, the quality of network visualization by different network embedding algorithms is evaluated visually. Fig.~\ref{nv} is an example by SDNE~\cite{wang2016kdd} where SDNE is applied to 20-NewsGroup. In
Fig.~\ref{nv}, each document is mapped into a two dimensional space as a point, and different colors on the
points represent the labels. As can be seen, network embedding preserves the intrinsic structure of the network, where
similar nodes are closer to each other than dissimilar nodes in the low-dimensional space. Also, LINE~\cite{tang2015line}, GraRep~\cite{cao2015grarep},
and EOE~\cite{xu2017embedding} are
applied to a citation network DBLP and generate meaningful layout of the network. Pan~\textit{et~al.}~\cite{pan2016tri} show the
visualization of another citation network Citeseer-M10~\cite{lim2016bibliographic} consisting of scientific publications from ten distinct research areas.

\subsection{Open Source Software}

\begin{table*}
\scriptsize
\caption{A summary of the source code}\label{asrw}
\centering
\begin{tabular}{c|c}
\hline
\multicolumn{2}{c}{\textbf{Structure and property preserving network embedding}} \\
\hline
Methods & Source code  \\
\hline
DeepWalk~\cite{perozzi2014deepwalk} & \url{https://github.com/phanein/deepwalk} \\
LINE~\cite{tang2015line} & \url{https://github.com/tangjianpku/LINE} \\
GraRep~\cite{cao2015grarep} & \url{https://github.com/ShelsonCao/GraRep} \\
SDNE~\cite{wang2016kdd} & \url{http://nrl.thumedia.org/structural-deep-network-embedding} \\
Node2vec~\cite{jure2016kdd} & \url{https://github.com/aditya-grover/node2vec} \\
DNGR~\cite{cao2016deep} & \url{https://github.com/ShelsonCao/DNGR} \\
M-NMF~\cite{wang2017community} & \url{http://nrl.thumedia.org/community-preserving-network-embedding} \\
GED~\cite{chen2017fast} & \url{https://users.ece.cmu.edu/~sihengc/publications.html} \\
Ou~\textit{et~al.}~\cite{ou2015non} & \url{http://nrl.thumedia.org/non-transitive-hashing-with-latent-similarity-components} \\
HOPE~\cite{ou2016kdd} & \url{http://nrl.thumedia.org/asymmetric-transitivity-preserving-graph-embedding} \\
\hline
\hline
\multicolumn{2}{c}{\textbf{Network embedding with side information}} \\
\hline
Methods & Source code  \\
\hline
MMDW~\cite{tu2016max} & \url{https://github.com/thunlp/mmdw} \\
TADW~\cite{yang2015network} & \url{https://github.com/thunlp/tadw} \\
TriDNR~\cite{pan2016tri} & \url{https://github.com/shiruipan/TriDNR} \\
\hline
\hline
\multicolumn{2}{c}{\textbf{Advanced information preserving network embedding}} \\
\hline
Methods & Source code  \\
\hline
Information diffusion~\cite{bourigault2014learning} & \url{https://github.com/ludc/social_network_diffusion_embeddings} \\
Cascade prediction~\cite{li2017deepcas} & \url{https://github.com/chengli-um/DeepCas} \\
Anomaly detection~\cite{hu2016embedding} & \url{https://github.com/hurenjun/EmbeddingAnomalyDetection} \\
Collaboration prediction~\cite{chen2017task} & \url{https://github.com/chentingpc/GuidedHeteEmbedding} \\
\hline
\end{tabular}
\label{scode}
\end{table*}

In Table~\ref{scode}, we provide a collection of links where one can find the source code of various network embedding methods.

\section{Conclusions and Future Research Directions}
\label{sec:future}

\begin{figure}
  \centering
  \includegraphics[width=3.0in]{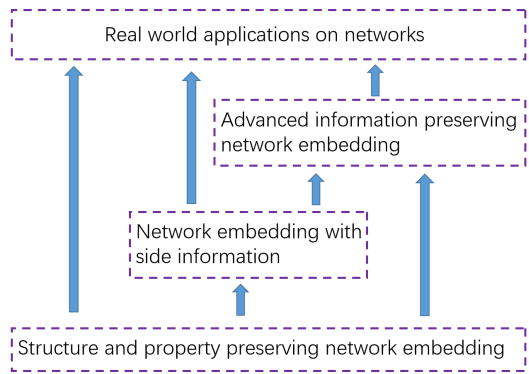}\\
  \caption{Relationship among different types of network embedding methods. }\label{fram}
\end{figure}

The above survey of the state-of-the-art network embedding algorithms clearly shows that it is still a young and promising research field. To apply network embedding to tackle practical applications, a frontmost question is to select the appropriate methods. In Fig.~\ref{fram} we show the relationship among different types of network embedding methods discussed in this survey.

The structure and property preserving network
embedding is the foundation. If one cannot
preserve well the network structure and retain the important network properties, in the embedding space serious information is loss, which hurts the analytic tasks in sequel. Based on
the structure and property preserving network embedding, one may apply the off-the-shelf machine learning
methods. If some side information is available, it can be incorporated
into network embedding. Furthermore, the domain knowledge of some certain applications as advanced information can be considered.


In the rest of this section, we discuss several interesting directions for future work.

\subsection{More Structures and Properties}

Although various methods are proposed to preserve structures and properties, such as first order and high order proximities, communities, asymmetric transitivity, and structural balance, due to the complexity of real world networks,
there are still some particular structures that are not fully considered in the existing network embedding methods. For example,
how to incorporate network motifs~\cite{benson2016higher}, one of the most common higher-order structures in a network, into network embedding remains an open problem.
Also, more complex local structures of a node can be considered to provide higher level constraints. The current assumption of network embedding is usually based
on the pairwise structure, that is, if two nodes have a link, then their representations are similar. This assumption can work well for some applications, such as
link prediction, but it cannot encode the centrality information of nodes, because the centrality of a node is usually related
to a more complex structure.
As another example, in several real world applications, an edge may involve more than two nodes, known as a hyperedge. Such a hypernetwork naturally
indicates richer relationships among nodes and has its own characteristics. Hypernetwork embedding
is important for some real applications.

The power law
distribution property indicates that most nodes in a network are associated with a small number of edges. Consequently, it is hard to learn an effective representation
for a node with limited information. How this property affects the performance of network embedding and how to improve
the embeddings of the minority nodes are still largely untouched.

\subsection{The Effect of Side Information} Section~\ref{sec:sideinfo} discusses a series of network embedding algorithms that preserve
side information in embedding. All the existing methods assume that there is an agreement between network structure and
side information. To what extent the assumption holds in real applications, however, remains an open question. The low correlation
of side information and structures may degrade the performance of network embedding.  Moreover, it is interesting to explore the complementarity between network structures and side information. More often than not, each information
may contain some knowledge that other information does not have.

Besides, in a heterogeneous information network, to measure the relevance of two objects, the meta path, a sequence of
object types with edge types in between, has been widely used. However, meta structure~\cite{huang2016meta}, which is essentially a directed acyclic graph
of object and edge types, provides a higher-order structure constraint. This suggests a huge potential direction for improving heterogeneous information network embedding.

\subsection{More Advanced Information and Tasks}
In general, most of network embedding algorithms are designed for general purposes, such as link prediction and node
classification. These network embedding methods mainly focus on general network structures and may not be specific to some target applications. Another important research direction is to explore the possibility of designing network
embedding for more specific applications. For example, whether network embedding is a new way to detect rumors in social
network~\cite{seo2012identifying,zhang2015automatic}? Can we use network embedding to infer social ties~\cite{tang2012inferring}?
Each real world application has its own characteristics, and incorporating their unique domain knowledge into network embedding
is a key.  The technical challenges here are how to model the specific domain knowledge as advanced information that can be integrated into network embedding in an effective manner.


\subsection{Dynamic Network Embedding}

Although many network embedding methods are proposed, they are mainly
designed for static networks. However, in real world applications, it is well recognized that many networks are evolving over time.
For example, in the Facebook network, friendships between users always dynamically change over time, e.g., new edges
are continuously added to the social network while some edges may be deleted. To learn the representations of nodes in a
dynamic network, the existing network embedding methods have to be run repeatedly for each time stamp, which is
very time consuming and may not meet the realtime processing demand. 
Most of the existing network embedding methods cannot be directly applied to large scale evolving networks. New network embedding algorithms, which
are able to tackle the dynamic nature of evolving networks, are highly desirable.


\subsection{More embedding spaces}

The existing network embedding methods embed a network into the Euclidean space.  In general, the principle of network embedding can be extended to other target spaces.  For example, recently some
studies~\cite{krioukov2010hyperbolic} assume that the underlying structure of a network is in the hyperbolic space. Under
this assumption, heterogeneous degree distributions and strong clustering emerge naturally, as they are the simple reflections of the negative curvature and metric property of the underlying hyperbolic geometry.
Exploring other embedding space is another interesting research direction.

\bibliographystyle{aaai}
\bibliography{wang}

\begin{thebibliography}{}

\bibitem[\protect\citeauthoryear{Adamic and Adar}{2003}]{adamic2003friends}
Adamic, L.~A., and Adar, E.
\newblock 2003.
\newblock Friends and neighbors on the web.
\newblock {\em Social networks} 25(3):211--230.

\bibitem[\protect\citeauthoryear{Akoglu, Tong, and
  Koutra}{2015}]{akoglu2015graph}
Akoglu, L.; Tong, H.; and Koutra, D.
\newblock 2015.
\newblock Graph based anomaly detection and description: a survey.
\newblock {\em Data Mining and Knowledge Discovery} 29(3):626--688.

\bibitem[\protect\citeauthoryear{Bayati \bgroup et al\mbox.\egroup
  }{2009}]{bayati2009algorithms}
Bayati, M.; Gerritsen, M.; Gleich, D.~F.; Saberi, A.; and Wang, Y.
\newblock 2009.
\newblock Algorithms for large, sparse network alignment problems.
\newblock In {\em Data Mining, 2009. ICDM'09. Ninth IEEE International
  Conference on},  705--710.
\newblock IEEE.

\bibitem[\protect\citeauthoryear{Belkin and Niyogi}{2002}]{belkin2002laplacian}
Belkin, M., and Niyogi, P.
\newblock 2002.
\newblock Laplacian eigenmaps and spectral techniques for embedding and
  clustering.
\newblock In {\em Advances in neural information processing systems},
  585--591.

\bibitem[\protect\citeauthoryear{Belkin and Niyogi}{2003}]{belkin2003laplacian}
Belkin, M., and Niyogi, P.
\newblock 2003.
\newblock Laplacian eigenmaps for dimensionality reduction and data
  representation.
\newblock {\em Neural computation} 15(6):1373--1396.

\bibitem[\protect\citeauthoryear{Benson, Gleich, and
  Leskovec}{2016}]{benson2016higher}
Benson, A.~R.; Gleich, D.~F.; and Leskovec, J.
\newblock 2016.
\newblock Higher-order organization of complex networks.
\newblock {\em Science} 353(6295):163--166.

\bibitem[\protect\citeauthoryear{Berline, Getzler, and
  Vergne}{2003}]{berline2003heat}
Berline, N.; Getzler, E.; and Vergne, M.
\newblock 2003.
\newblock {\em Heat kernels and Dirac operators}.
\newblock Springer Science \& Business Media.

\bibitem[\protect\citeauthoryear{Bollacker \bgroup et al\mbox.\egroup
  }{2008}]{bollacker2008freebase}
Bollacker, K.; Evans, C.; Paritosh, P.; Sturge, T.; and Taylor, J.
\newblock 2008.
\newblock Freebase: a collaboratively created graph database for structuring
  human knowledge.
\newblock In {\em Proceedings of the 2008 ACM SIGMOD international conference
  on Management of data},  1247--1250.
\newblock AcM.

\bibitem[\protect\citeauthoryear{Bourigault \bgroup et al\mbox.\egroup
  }{2014}]{bourigault2014learning}
Bourigault, S.; Lagnier, C.; Lamprier, S.; Denoyer, L.; and Gallinari, P.
\newblock 2014.
\newblock Learning social network embeddings for predicting information
  diffusion.
\newblock In {\em Proceedings of the 7th ACM international conference on Web
  search and data mining},  393--402.
\newblock ACM.

\bibitem[\protect\citeauthoryear{Breitkreutz \bgroup et al\mbox.\egroup
  }{2007}]{breitkreutz2007biogrid}
Breitkreutz, B.-J.; Stark, C.; Reguly, T.; Boucher, L.; Breitkreutz, A.;
  Livstone, M.; Oughtred, R.; Lackner, D.~H.; B{\"a}hler, J.; Wood, V.; et~al.
\newblock 2007.
\newblock The biogrid interaction database: 2008 update.
\newblock {\em Nucleic acids research} 36(suppl\_1):D637--D640.

\bibitem[\protect\citeauthoryear{Burt}{2004}]{burt2004structural}
Burt, R.~S.
\newblock 2004.
\newblock Structural holes and good ideas.
\newblock {\em American journal of sociology} 110(2):349--399.

\bibitem[\protect\citeauthoryear{Cai \bgroup et al\mbox.\egroup
  }{2011}]{cai2011graph}
Cai, D.; He, X.; Han, J.; and Huang, T.~S.
\newblock 2011.
\newblock Graph regularized nonnegative matrix factorization for data
  representation.
\newblock {\em IEEE Transactions on Pattern Analysis and Machine Intelligence}
  33(8):1548--1560.

\bibitem[\protect\citeauthoryear{Cao, Lu, and Xu}{2015}]{cao2015grarep}
Cao, S.; Lu, W.; and Xu, Q.
\newblock 2015.
\newblock Grarep: Learning graph representations with global structural
  information.
\newblock In {\em Proceedings of the 24th ACM International on Conference on
  Information and Knowledge Management},  891--900.
\newblock ACM.

\bibitem[\protect\citeauthoryear{Cao, Lu, and Xu}{2016}]{cao2016deep}
Cao, S.; Lu, W.; and Xu, Q.
\newblock 2016.
\newblock Deep neural networks for learning graph representations.
\newblock In {\em Proceedings of the Thirtieth AAAI Conference on Artificial
  Intelligence},  1145--1152.
\newblock AAAI Press.

\bibitem[\protect\citeauthoryear{Cartwright and
  Harary}{1956}]{cartwright1956structural}
Cartwright, D., and Harary, F.
\newblock 1956.
\newblock Structural balance: a generalization of heider's theory.
\newblock {\em Psychological review} 63(5):277.

\bibitem[\protect\citeauthoryear{Chang and Blei}{2009}]{chang2009relational}
Chang, J., and Blei, D.~M.
\newblock 2009.
\newblock Relational topic models for document networks.
\newblock In {\em International conference on artificial intelligence and
  statistics},  81--88.

\bibitem[\protect\citeauthoryear{Chang \bgroup et al\mbox.\egroup
  }{2015}]{chang2015heterogeneous}
Chang, S.; Han, W.; Tang, J.; Qi, G.-J.; Aggarwal, C.~C.; and Huang, T.~S.
\newblock 2015.
\newblock Heterogeneous network embedding via deep architectures.
\newblock In {\em Proceedings of the 21th ACM SIGKDD International Conference
  on Knowledge Discovery and Data Mining},  119--128.
\newblock ACM.

\bibitem[\protect\citeauthoryear{Chen and Sun}{2017}]{chen2017task}
Chen, T., and Sun, Y.
\newblock 2017.
\newblock Task-guided and path-augmented heterogeneous network embedding for
  author identification.
\newblock In {\em Proceedings of the Tenth ACM International Conference on Web
  Search and Data Mining},  295--304.
\newblock ACM.

\bibitem[\protect\citeauthoryear{Chen \bgroup et al\mbox.\egroup
  }{2017}]{chen2017fast}
Chen, S.; Niu, S.; Akoglu, L.; Kova{\v{c}}evi{\'c}, J.; and Faloutsos, C.
\newblock 2017.
\newblock Fast, warped graph embedding: Unifying framework and one-click
  algorithm.
\newblock {\em arXiv preprint arXiv:1702.05764}.

\bibitem[\protect\citeauthoryear{Chua \bgroup et al\mbox.\egroup
  }{2009}]{chua2009nus}
Chua, T.-S.; Tang, J.; Hong, R.; Li, H.; Luo, Z.; and Zheng, Y.
\newblock 2009.
\newblock Nus-wide: a real-world web image database from national university of
  singapore.
\newblock In {\em Proceedings of the ACM international conference on image and
  video retrieval}, ~48.
\newblock ACM.

\bibitem[\protect\citeauthoryear{Chung}{1997}]{chung1997spectral}
Chung, F.~R.
\newblock 1997.
\newblock {\em Spectral graph theory}.
\newblock Number~92. American Mathematical Soc.

\bibitem[\protect\citeauthoryear{Cygan \bgroup et al\mbox.\egroup
  }{2015}]{cygan2015sitting}
Cygan, M.; Pilipczuk, M.; Pilipczuk, M.; and Wojtaszczyk, J.~O.
\newblock 2015.
\newblock Sitting closer to friends than enemies, revisited.
\newblock {\em Theory of computing systems} 56(2):394--405.

\bibitem[\protect\citeauthoryear{De~Choudhury \bgroup et al\mbox.\egroup
  }{2010}]{de2010does}
De~Choudhury, M.; Lin, Y.-R.; Sundaram, H.; Candan, K.~S.; Xie, L.; Kelliher,
  A.; et~al.
\newblock 2010.
\newblock How does the data sampling strategy impact the discovery of
  information diffusion in social media?
\newblock {\em ICWSM} 10:34--41.

\bibitem[\protect\citeauthoryear{Fan \bgroup et al\mbox.\egroup
  }{2008}]{fan2008liblinear}
Fan, R.-E.; Chang, K.-W.; Hsieh, C.-J.; Wang, X.-R.; and Lin, C.-J.
\newblock 2008.
\newblock Liblinear: A library for large linear classification.
\newblock {\em Journal of machine learning research} 9(Aug):1871--1874.

\bibitem[\protect\citeauthoryear{Fu and Ma}{2012}]{fu2012graph}
Fu, Y., and Ma, Y.
\newblock 2012.
\newblock {\em Graph embedding for pattern analysis}.
\newblock Springer Science \&amp; Business Media.

\bibitem[\protect\citeauthoryear{Getoor and Diehl}{2005}]{getoor2005link}
Getoor, L., and Diehl, C.~P.
\newblock 2005.
\newblock Link mining: a survey.
\newblock {\em Acm Sigkdd Explorations Newsletter} 7(2):3--12.

\bibitem[\protect\citeauthoryear{Girvan and Newman}{2002}]{girvan2002community}
Girvan, M., and Newman, M.~E.
\newblock 2002.
\newblock Community structure in social and biological networks.
\newblock {\em Proceedings of the national academy of sciences}
  99(12):7821--7826.

\bibitem[\protect\citeauthoryear{Grover and Leskovec}{2016}]{jure2016kdd}
Grover, A., and Leskovec, J.
\newblock 2016.
\newblock node2vec: Scalable feature learning for networks.
\newblock In {\em Proceedings of the 22nd ACM SIGKDD international conference
  on Knowledge discovery and data mining},  1225--1234.
\newblock ACM.

\bibitem[\protect\citeauthoryear{Guille \bgroup et al\mbox.\egroup
  }{2013}]{guille2013information}
Guille, A.; Hacid, H.; Favre, C.; and Zighed, D.~A.
\newblock 2013.
\newblock Information diffusion in online social networks: A survey.
\newblock {\em ACM Sigmod Record} 42(2):17--28.

\bibitem[\protect\citeauthoryear{He and Niyogi}{2004}]{he2004locality}
He, X., and Niyogi, P.
\newblock 2004.
\newblock Locality preserving projections.
\newblock In {\em Advances in neural information processing systems},
  153--160.

\bibitem[\protect\citeauthoryear{Hearst \bgroup et al\mbox.\egroup
  }{1998}]{hearst1998support}
Hearst, M.~A.; Dumais, S.~T.; Osuna, E.; Platt, J.; and Scholkopf, B.
\newblock 1998.
\newblock Support vector machines.
\newblock {\em IEEE Intelligent Systems and their Applications} 13(4):18--28.

\bibitem[\protect\citeauthoryear{Herman, Melan{\c{c}}on, and
  Marshall}{2000}]{herman2000graph}
Herman, I.; Melan{\c{c}}on, G.; and Marshall, M.~S.
\newblock 2000.
\newblock Graph visualization and navigation in information visualization: A
  survey.
\newblock {\em IEEE Transactions on visualization and computer graphics}
  6(1):24--43.

\bibitem[\protect\citeauthoryear{Hochreiter and
  Schmidhuber}{1997}]{hochreiter1997long}
Hochreiter, S., and Schmidhuber, J.
\newblock 1997.
\newblock Long short-term memory.
\newblock {\em Neural computation} 9(8):1735--1780.

\bibitem[\protect\citeauthoryear{Hu \bgroup et al\mbox.\egroup
  }{2016}]{hu2016embedding}
Hu, R.; Aggarwal, C.~C.; Ma, S.; and Huai, J.
\newblock 2016.
\newblock An embedding approach to anomaly detection.
\newblock In {\em Data Engineering (ICDE), 2016 IEEE 32nd International
  Conference on},  385--396.
\newblock IEEE.

\bibitem[\protect\citeauthoryear{Huang and
  Mamoulis}{2017}]{huang2017heterogeneous}
Huang, Z., and Mamoulis, N.
\newblock 2017.
\newblock Heterogeneous information network embedding for meta path based
  proximity.
\newblock {\em arXiv preprint arXiv:1701.05291}.

\bibitem[\protect\citeauthoryear{Huang \bgroup et al\mbox.\egroup
  }{2014}]{huang2014mining}
Huang, H.; Tang, J.; Wu, S.; Liu, L.; et~al.
\newblock 2014.
\newblock Mining triadic closure patterns in social networks.
\newblock In {\em Proceedings of the 23rd international conference on World
  wide web},  499--504.
\newblock ACM.

\bibitem[\protect\citeauthoryear{Huang \bgroup et al\mbox.\egroup
  }{2016}]{huang2016meta}
Huang, Z.; Zheng, Y.; Cheng, R.; Sun, Y.; Mamoulis, N.; and Li, X.
\newblock 2016.
\newblock Meta structure: Computing relevance in large heterogeneous
  information networks.
\newblock In {\em Proceedings of the 22nd ACM SIGKDD International Conference
  on Knowledge Discovery and Data Mining},  1595--1604.
\newblock ACM.

\bibitem[\protect\citeauthoryear{Huang, Li, and Hu}{2017}]{huang2017label}
Huang, X.; Li, J.; and Hu, X.
\newblock 2017.
\newblock Label informed attributed network embedding.
\newblock In {\em Proceedings of 10th ACM International Conference on Web
  Search and Data Mining (WSDM)}.

\bibitem[\protect\citeauthoryear{Jacob, Denoyer, and
  Gallinari}{2014}]{jacob2014learning}
Jacob, Y.; Denoyer, L.; and Gallinari, P.
\newblock 2014.
\newblock Learning latent representations of nodes for classifying in
  heterogeneous social networks.
\newblock In {\em Proceedings of the 7th ACM international conference on Web
  search and data mining},  373--382.
\newblock ACM.

\bibitem[\protect\citeauthoryear{Katz}{1953}]{katz1953new}
Katz, L.
\newblock 1953.
\newblock A new status index derived from sociometric analysis.
\newblock {\em Psychometrika} 18(1):39--43.

\bibitem[\protect\citeauthoryear{Krioukov \bgroup et al\mbox.\egroup
  }{2010}]{krioukov2010hyperbolic}
Krioukov, D.; Papadopoulos, F.; Kitsak, M.; Vahdat, A.; and Bogun{\'a}, M.
\newblock 2010.
\newblock Hyperbolic geometry of complex networks.
\newblock {\em Physical Review E} 82(3):036106.

\bibitem[\protect\citeauthoryear{Le and Lauw}{2014}]{le2014probabilistic}
Le, T.~M., and Lauw, H.~W.
\newblock 2014.
\newblock Probabilistic latent document network embedding.
\newblock In {\em Data Mining (ICDM), 2014 IEEE International Conference on},
  270--279.
\newblock IEEE.

\bibitem[\protect\citeauthoryear{Lee and Seung}{2001}]{lee2001algorithms}
Lee, D.~D., and Seung, H.~S.
\newblock 2001.
\newblock Algorithms for non-negative matrix factorization.
\newblock In {\em Advances in neural information processing systems},
  556--562.

\bibitem[\protect\citeauthoryear{Leskovec and Krevl}{2016}]{leskovec2016snap}
Leskovec, J., and Krevl, A.
\newblock 2016.
\newblock Snap datasets: Stanford large network dataset collection (2014).
\newblock {\em URL http://snap. stanford. edu/data}.

\bibitem[\protect\citeauthoryear{Leskovec, Kleinberg, and
  Faloutsos}{2007}]{leskovec2007graph}
Leskovec, J.; Kleinberg, J.; and Faloutsos, C.
\newblock 2007.
\newblock Graph evolution: Densification and shrinking diameters.
\newblock {\em ACM Transactions on Knowledge Discovery from Data (TKDD)}
  1(1):2.

\bibitem[\protect\citeauthoryear{Levy and Goldberg}{2014}]{levy2014neural}
Levy, O., and Goldberg, Y.
\newblock 2014.
\newblock Neural word embedding as implicit matrix factorization.
\newblock In {\em Advances in neural information processing systems},
  2177--2185.

\bibitem[\protect\citeauthoryear{Li \bgroup et al\mbox.\egroup
  }{2017}]{li2017deepcas}
Li, C.; Ma, J.; Guo, X.; and Mei, Q.
\newblock 2017.
\newblock Deepcas: an end-to-end predictor of information cascades.
\newblock In {\em Proceedings of the 26th International Conference on World
  Wide Web},  577--586.
\newblock International World Wide Web Conferences Steering Committee.

\bibitem[\protect\citeauthoryear{Liben-Nowell and
  Kleinberg}{2007}]{liben2007link}
Liben-Nowell, D., and Kleinberg, J.
\newblock 2007.
\newblock The link-prediction problem for social networks.
\newblock {\em journal of the Association for Information Science and
  Technology} 58(7):1019--1031.

\bibitem[\protect\citeauthoryear{Lim and Buntine}{2016}]{lim2016bibliographic}
Lim, K.~W., and Buntine, W.
\newblock 2016.
\newblock Bibliographic analysis with the citation network topic model.
\newblock {\em arXiv preprint arXiv:1609.06826}.

\bibitem[\protect\citeauthoryear{Lov{\'a}sz and
  Plummer}{2009}]{lovasz2009matching}
Lov{\'a}sz, L., and Plummer, M.~D.
\newblock 2009.
\newblock {\em Matching theory}, volume 367.
\newblock American Mathematical Soc.

\bibitem[\protect\citeauthoryear{L{\"u} and Zhou}{2011}]{lu2011link}
L{\"u}, L., and Zhou, T.
\newblock 2011.
\newblock Link prediction in complex networks: A survey.
\newblock {\em Physica A: statistical mechanics and its applications}
  390(6):1150--1170.

\bibitem[\protect\citeauthoryear{Maaten and
  Hinton}{2008}]{maaten2008visualizing}
Maaten, L. v.~d., and Hinton, G.
\newblock 2008.
\newblock Visualizing data using t-sne.
\newblock {\em Journal of Machine Learning Research} 9(Nov):2579--2605.

\bibitem[\protect\citeauthoryear{MacQueen and others}{1967}]{macqueen1967some}
MacQueen, J., et~al.
\newblock 1967.
\newblock Some methods for classification and analysis of multivariate
  observations.
\newblock In {\em Proceedings of the fifth Berkeley symposium on mathematical
  statistics and probability}, volume~1,  281--297.
\newblock Oakland, CA, USA.

\bibitem[\protect\citeauthoryear{Mahoney}{2011}]{mahoney2011large}
Mahoney, M.
\newblock 2011.
\newblock Large text compression benchmark.

\bibitem[\protect\citeauthoryear{Man \bgroup et al\mbox.\egroup
  }{2016}]{man2016predict}
Man, T.; Shen, H.; Liu, S.; Jin, X.; and Cheng, X.
\newblock 2016.
\newblock Predict anchor links across social networks via an embedding
  approach.
\newblock IJCAI.

\bibitem[\protect\citeauthoryear{McCallum \bgroup et al\mbox.\egroup
  }{2000}]{mccallum2000automating}
McCallum, A.~K.; Nigam, K.; Rennie, J.; and Seymore, K.
\newblock 2000.
\newblock Automating the construction of internet portals with machine
  learning.
\newblock {\em Information Retrieval} 3(2):127--163.

\bibitem[\protect\citeauthoryear{Mikolov \bgroup et al\mbox.\egroup
  }{2010}]{mikolov2010recurrent}
Mikolov, T.; Karafi{\'a}t, M.; Burget, L.; Cernock{\`y}, J.; and Khudanpur, S.
\newblock 2010.
\newblock Recurrent neural network based language model.
\newblock In {\em Interspeech}, volume~2, ~3.

\bibitem[\protect\citeauthoryear{Mikolov \bgroup et al\mbox.\egroup
  }{2013a}]{mikolov2013efficient}
Mikolov, T.; Chen, K.; Corrado, G.; and Dean, J.
\newblock 2013a.
\newblock Efficient estimation of word representations in vector space.
\newblock {\em arXiv preprint arXiv:1301.3781}.

\bibitem[\protect\citeauthoryear{Mikolov \bgroup et al\mbox.\egroup
  }{2013b}]{mikolov2013distributed}
Mikolov, T.; Sutskever, I.; Chen, K.; Corrado, G.~S.; and Dean, J.
\newblock 2013b.
\newblock Distributed representations of words and phrases and their
  compositionality.
\newblock In {\em Advances in neural information processing systems},
  3111--3119.

\bibitem[\protect\citeauthoryear{Natarajan and
  Dhillon}{2014}]{natarajan2014inductive}
Natarajan, N., and Dhillon, I.~S.
\newblock 2014.
\newblock Inductive matrix completion for predicting gene--disease
  associations.
\newblock {\em Bioinformatics} 30(12):i60--i68.

\bibitem[\protect\citeauthoryear{Newman}{2006}]{newman2006finding}
Newman, M.~E.
\newblock 2006.
\newblock Finding community structure in networks using the eigenvectors of
  matrices.
\newblock {\em Physical review E} 74(3):036104.

\bibitem[\protect\citeauthoryear{Ou \bgroup et al\mbox.\egroup
  }{2015}]{ou2015non}
Ou, M.; Cui, P.; Wang, F.; Wang, J.; and Zhu, W.
\newblock 2015.
\newblock Non-transitive hashing with latent similarity components.
\newblock In {\em Proceedings of the 21th ACM SIGKDD International Conference
  on Knowledge Discovery and Data Mining},  895--904.
\newblock ACM.

\bibitem[\protect\citeauthoryear{Ou \bgroup et al\mbox.\egroup
  }{2016}]{ou2016kdd}
Ou, M.; Cui, P.; Pei, J.; Zhang, Z.; and Zhu, W.
\newblock 2016.
\newblock Asymmetric transitivity preserving graph embedding.
\newblock In {\em Proceedings of the 22nd ACM SIGKDD international conference
  on Knowledge discovery and data mining},  672--681.
\newblock ACM.

\bibitem[\protect\citeauthoryear{Paige and Saunders}{1981}]{paige1981towards}
Paige, C.~C., and Saunders, M.~A.
\newblock 1981.
\newblock Towards a generalized singular value decomposition.
\newblock {\em SIAM Journal on Numerical Analysis} 18(3):398--405.

\bibitem[\protect\citeauthoryear{Pan \bgroup et al\mbox.\egroup
  }{2016}]{pan2016tri}
Pan, S.; Wu, J.; Zhu, X.; Zhang, C.; and Wang, Y.
\newblock 2016.
\newblock Tri-party deep network representation.
\newblock {\em Network} 11(9):12.

\bibitem[\protect\citeauthoryear{Perozzi, Al-Rfou, and
  Skiena}{2014}]{perozzi2014deepwalk}
Perozzi, B.; Al-Rfou, R.; and Skiena, S.
\newblock 2014.
\newblock Deepwalk: Online learning of social representations.
\newblock In {\em Proceedings of the 20th ACM SIGKDD international conference
  on Knowledge discovery and data mining},  701--710.
\newblock ACM.

\bibitem[\protect\citeauthoryear{Roweis and Saul}{2000}]{roweis2000nonlinear}
Roweis, S.~T., and Saul, L.~K.
\newblock 2000.
\newblock Nonlinear dimensionality reduction by locally linear embedding.
\newblock {\em science} 290(5500):2323--2326.

\bibitem[\protect\citeauthoryear{Ruck \bgroup et al\mbox.\egroup
  }{1990}]{ruck1990multilayer}
Ruck, D.~W.; Rogers, S.~K.; Kabrisky, M.; Oxley, M.~E.; and Suter, B.~W.
\newblock 1990.
\newblock The multilayer perceptron as an approximation to a bayes optimal
  discriminant function.
\newblock {\em IEEE Transactions on Neural Networks} 1(4):296--298.

\bibitem[\protect\citeauthoryear{Sen \bgroup et al\mbox.\egroup
  }{2008}]{sen2008collective}
Sen, P.; Namata, G.; Bilgic, M.; Getoor, L.; Galligher, B.; and Eliassi-Rad, T.
\newblock 2008.
\newblock Collective classification in network data.
\newblock {\em AI magazine} 29(3):93.

\bibitem[\protect\citeauthoryear{Seo, Mohapatra, and
  Abdelzaher}{2012}]{seo2012identifying}
Seo, E.; Mohapatra, P.; and Abdelzaher, T.
\newblock 2012.
\newblock Identifying rumors and their sources in social networks.
\newblock {\em SPIE defense, security, and sensing}  83891I--83891I.

\bibitem[\protect\citeauthoryear{Staudt, Sazonovs, and
  Meyerhenke}{}]{staudtnetworkit}
Staudt, C.; Sazonovs, A.; and Meyerhenke, H.
\newblock Networkit: A tool suite for large-scale network analysis.
\newblock {\em Network Science To appear}.

\bibitem[\protect\citeauthoryear{Sun \bgroup et al\mbox.\egroup
  }{2011}]{sun2011pathsim}
Sun, Y.; Han, J.; Yan, X.; Yu, P.~S.; and Wu, T.
\newblock 2011.
\newblock Pathsim: Meta path-based top-k similarity search in heterogeneous
  information networks.
\newblock {\em Proceedings of the VLDB Endowment} 4(11):992--1003.

\bibitem[\protect\citeauthoryear{Sun \bgroup et al\mbox.\egroup
  }{2016}]{sun2016general}
Sun, X.; Guo, J.; Ding, X.; and Liu, T.
\newblock 2016.
\newblock A general framework for content-enhanced network representation
  learning.
\newblock {\em arXiv preprint arXiv:1610.02906}.

\bibitem[\protect\citeauthoryear{Tang and Liu}{2009a}]{tang2009relational}
Tang, L., and Liu, H.
\newblock 2009a.
\newblock Relational learning via latent social dimensions.
\newblock In {\em Proceedings of the 15th ACM SIGKDD international conference
  on Knowledge discovery and data mining},  817--826.
\newblock ACM.

\bibitem[\protect\citeauthoryear{tang and Liu}{2009b}]{tang2009scalable}
tang, L., and Liu, H.
\newblock 2009b.
\newblock Scalable learning of collective behavior based on sparse social
  dimensions.
\newblock In {\em Proceedings of the 18th ACM conference on Information and
  knowledge management},  1107--1116.
\newblock ACM.

\bibitem[\protect\citeauthoryear{Tang \bgroup et al\mbox.\egroup
  }{2008}]{tang2008arnetminer}
Tang, J.; Zhang, J.; Yao, L.; Li, J.; Zhang, L.; and Su, Z.
\newblock 2008.
\newblock Arnetminer: extraction and mining of academic social networks.
\newblock In {\em Proceedings of the 14th ACM SIGKDD international conference
  on Knowledge discovery and data mining},  990--998.
\newblock ACM.

\bibitem[\protect\citeauthoryear{Tang \bgroup et al\mbox.\egroup
  }{2015}]{tang2015line}
Tang, J.; Qu, M.; Wang, M.; Zhang, M.; Yan, J.; and Mei, Q.
\newblock 2015.
\newblock Line: Large-scale information network embedding.
\newblock In {\em Proceedings of the 24th International Conference on World
  Wide Web},  1067--1077.
\newblock ACM.

\bibitem[\protect\citeauthoryear{Tang, Lou, and
  Kleinberg}{2012}]{tang2012inferring}
Tang, J.; Lou, T.; and Kleinberg, J.
\newblock 2012.
\newblock Inferring social ties across heterogenous networks.
\newblock In {\em Proceedings of the fifth ACM international conference on Web
  search and data mining},  743--752.
\newblock ACM.

\bibitem[\protect\citeauthoryear{Tenenbaum, De~Silva, and
  Langford}{2000}]{tenenbaum2000global}
Tenenbaum, J.~B.; De~Silva, V.; and Langford, J.~C.
\newblock 2000.
\newblock A global geometric framework for nonlinear dimensionality reduction.
\newblock {\em science} 290(5500):2319--2323.

\bibitem[\protect\citeauthoryear{Tian \bgroup et al\mbox.\egroup
  }{2014}]{tian2014learning}
Tian, F.; Gao, B.; Cui, Q.; Chen, E.; and Liu, T.-Y.
\newblock 2014.
\newblock Learning deep representations for graph clustering.
\newblock In {\em AAAI},  1293--1299.

\bibitem[\protect\citeauthoryear{Traud, Mucha, and
  Porter}{2012}]{traud2012social}
Traud, A.~L.; Mucha, P.~J.; and Porter, M.~A.
\newblock 2012.
\newblock Social structure of facebook networks.
\newblock {\em Physica A: Statistical Mechanics and its Applications}
  391(16):4165--4180.

\bibitem[\protect\citeauthoryear{Tu \bgroup et al\mbox.\egroup
  }{2016}]{tu2016max}
Tu, C.; Zhang, W.; Liu, Z.; and Sun, M.
\newblock 2016.
\newblock Max-margin deepwalk: discriminative learning of network
  representation.
\newblock In {\em Proceedings of the Twenty-Fifth International Joint
  Conference on Artificial Intelligence (IJCAI 2016)},  3889--3895.

\bibitem[\protect\citeauthoryear{Vincent \bgroup et al\mbox.\egroup
  }{2010}]{vincent2010stacked}
Vincent, P.; Larochelle, H.; Lajoie, I.; Bengio, Y.; and Manzagol, P.-A.
\newblock 2010.
\newblock Stacked denoising autoencoders: Learning useful representations in a
  deep network with a local denoising criterion.
\newblock {\em Journal of Machine Learning Research} 11(Dec):3371--3408.

\bibitem[\protect\citeauthoryear{Wang \bgroup et al\mbox.\egroup
  }{2017a}]{wang2017signed}
Wang, S.; Tang, J.; Aggarwal, C.; Chang, Y.; and Liu, H.
\newblock 2017a.
\newblock Signed network embedding in social media.
\newblock In {\em Proceedings of the 2017 SIAM International Conference on Data
  Mining},  327--335.
\newblock SIAM.

\bibitem[\protect\citeauthoryear{Wang \bgroup et al\mbox.\egroup
  }{2017b}]{wang2017community}
Wang, X.; Cui, P.; Wang, J.; Pei, J.; Zhu, W.; and Yang, S.
\newblock 2017b.
\newblock Community preserving network embedding.

\bibitem[\protect\citeauthoryear{Wang, Cui, and Zhu}{2016}]{wang2016kdd}
Wang, D.; Cui, P.; and Zhu, W.
\newblock 2016.
\newblock Structural deep network embedding.
\newblock In {\em Proceedings of the 22nd ACM SIGKDD international conference
  on Knowledge discovery and data mining},  1225--1234.
\newblock ACM.

\bibitem[\protect\citeauthoryear{Xu \bgroup et al\mbox.\egroup
  }{2017}]{xu2017embedding}
Xu, L.; Wei, X.; Cao, J.; and Yu, P.~S.
\newblock 2017.
\newblock Embedding of embedding (eoe): Joint embedding for coupled
  heterogeneous networks.
\newblock In {\em Proceedings of the Tenth ACM International Conference on Web
  Search and Data Mining},  741--749.
\newblock ACM.

\bibitem[\protect\citeauthoryear{Yan \bgroup et al\mbox.\egroup
  }{2005}]{yan2005graph}
Yan, S.; Xu, D.; Zhang, B.; and Zhang, H.-J.
\newblock 2005.
\newblock Graph embedding: A general framework for dimensionality reduction.
\newblock In {\em Computer Vision and Pattern Recognition, 2005. CVPR 2005.
  IEEE Computer Society Conference on}, volume~2,  830--837.
\newblock IEEE.

\bibitem[\protect\citeauthoryear{Yang \bgroup et al\mbox.\egroup
  }{2015}]{yang2015network}
Yang, C.; Liu, Z.; Zhao, D.; Sun, M.; and Chang, E.~Y.
\newblock 2015.
\newblock Network representation learning with rich text information.
\newblock In {\em Proceedings of the 24th International Joint Conference on
  Artificial Intelligence, Buenos Aires, Argentina},  2111--2117.

\bibitem[\protect\citeauthoryear{Yang \bgroup et al\mbox.\egroup
  }{2017}]{yang2017end}
Yang, X.; Chen, Y.-N.; Hakkani-T{\"u}r, D.; Crook, P.; Li, X.; Gao, J.; and
  Deng, L.
\newblock 2017.
\newblock End-to-end joint learning of natural language understanding and
  dialogue manager.
\newblock In {\em Acoustics, Speech and Signal Processing (ICASSP), 2017 IEEE
  International Conference on},  5690--5694.
\newblock IEEE.

\bibitem[\protect\citeauthoryear{Yeung \bgroup et al\mbox.\egroup
  }{2016}]{yeung2016end}
Yeung, S.; Russakovsky, O.; Mori, G.; and Fei-Fei, L.
\newblock 2016.
\newblock End-to-end learning of action detection from frame glimpses in
  videos.
\newblock In {\em Proceedings of the IEEE Conference on Computer Vision and
  Pattern Recognition},  2678--2687.

\bibitem[\protect\citeauthoryear{Zhang \bgroup et al\mbox.\egroup
  }{2015}]{zhang2015automatic}
Zhang, Q.; Zhang, S.; Dong, J.; Xiong, J.; and Cheng, X.
\newblock 2015.
\newblock Automatic detection of rumor on social network.
\newblock In {\em Natural Language Processing and Chinese Computing}. Springer.
\newblock  113--122.

\end{thebibliography}

\end{document}